\documentclass[aps, prd, preprint, amsfonts, floatfix,nofootinbib]{revtex4}

\usepackage{amsfonts}
\usepackage{amsmath}
\usepackage{color}
\usepackage{hyperref}

\usepackage{pict2e}
\usepackage{graphicx}

\DeclareMathOperator{\diag}{diag}
\newcommand{\be}{\begin{eqnarray}}
\newcommand{\ee}{\end{eqnarray}}

\newcommand{\nn}{\nonumber}
\newcommand{\Tr }{{\rm Tr}}

\newcommand{\mat}{\left ( \begin{array}{cc}}
	\newcommand{\emat}{\end{array} \right )}
\newcommand{\vect}{\left ( \begin{array}{c}}
	\newcommand{\evect}{\end{array} \right )}

\newcommand{\hc}{^\dagger}
\newcommand{\inv}{^{-1}}

\newcommand{\eins}{\leavevmode\hbox{\small1\kern-3.8pt\normalsize1}}
\newcommand{\CondS}{(\ref{Eq:CondTrace0}-\ref{Eq:strength})}
\newcommand{\EnsAntiRand}{{\it Ensemble 1}}
\newcommand{\EnsAntiNone}{{\it Ensemble 2}}
\newcommand{\EnsAntiGam}{{\it Ensemble 3}}
\newcommand{\EnsCh}{{\it Ensemble 4}}

\begin{document}
\title[Universal Broadening of Zero Modes]
 {\bf{Universal Broadening of Zero Modes:}\\ A General Framework and Identification}

\author{M. Kieburg}

\author{A. Mielke}
\affiliation{Faculty of Physics, Bielefeld University,  P.O. Box 100131, D-33501 Bielefeld, Germany}

\author{K. Splittorff}
\affiliation{Niels Bohr Institute, University of Copenhagen, Blegdamsvej 17, 2100 Copenhagen, Denmark}

\begin{abstract}
We consider the smallest eigenvalues of perturbed Hermitian operators with zero modes, either topological or system specific. To leading order for small generic perturbation we show that the corresponding eigenvalues broaden to a Gaussian random matrix ensemble of size $\nu\times\nu$, where $\nu$ is the number of zero modes.
This observation unifies and extends a number of results within chiral random matrix theory and effective field theory and clarifies under which conditions they apply. The scaling of the former zero modes with the volume differs from the eigenvalues in the bulk, which we propose as an indicator to identify them in experiments.
These results hold for all ten symmetric spaces in the Altland-Zirnbauer classification and build on two facts. Firstly, the broadened zero modes decouple from the bulk eigenvalues and secondly, the mixing from eigenstates of the perturbation form a Central Limit Theorem argument for matrices.
\end{abstract}


\date{\today}
\maketitle
\newpage

\section{Introduction} \label{Sec:Introduction}

When studying the local (microscopic) spectral statistics of eigenvalues of operators, random matrix theory (RMT) provides universal results, see e.g.~\cite{Mehta,GMG,book} and references therein. One particular intriguing regime of eigenvalues is that close to the origin or at a spectral gap. These eigenvalues hold information about the large scale properties of the underlying system, because they are of the order of inverse system size. For instance, analysis of Dirac eigenvalues close to the origin has lead to a greater understanding of chiral symmetry breaking in QCD~\cite{VerbZahed,VerbaarschotThreeFold,JacBeta2,DOTV}.

The form of RMT relevant for a given physical system depends on the symmetries of the system. Not only the pure symmetry classes have been of interest, see~\cite{Mehta,book,Dyson,Martin,Casell,BernardLeClair,Magnea} for symmetry classifications in RMT and~\cite{VerbaarschotThreeFold,Dyson,Hueffmann,Casell2,AlexMartin,Ludwig,DFI,Slager2,MarioJac,Chiu,Slager1,MarioTim} for the classification of these symmetries in physical systems. It has been necessary to extend the random matrix models to two-matrix models, see e.g.~\cite{PandeyMehta,MehtaPandey,FHN,NF99,KTNK,KatoriTanemura,AN,MarioTakuya,AKMV} or even many matrix models like products and sums, e.g.~\cite{BougerolLacroix,QiuWicks,Kumar,AkemannIpsen,ACK,Mario2017} and references therein. Those models describe transitions between different symmetry classes. These are needed because no realistic system is completely pure, but usually perceives perturbations from its environment.

Degeneracies are vulnerable to perturbations which violate the condition that caused the degeneracy. For example topological zero modes are broadened due to residual interactions that break topology. This broadening can be used as a measure of the perturbation strength~\cite{DelDebbio:2005qa,DWW2011,DHS2012,KieburgWilson,CGRSZ}.
Topological modes are relevant in both high energy physics~\cite{VerbZahed,VerbaarschotThreeFold,JacBeta2,DOTV,LeutSmil,RMT_2_EFT-1,RMT_2_EFT-2,ADMNUniversality,Srednicki} and condensed matter systems~\cite{Chiu,Ivanov,Kitaev,HK,BagretsAltland,BeenakkerMajorana,Wilczek,BeenakkerRMT,Elliot}. For solid state physics, interactions in many-body systems perturbed by thermal fluctuations of the kind found in topological superconductors has been proposed to broaden zero modes~\cite{Kitaev,HK,BeenakkerMajorana,Wilczek,Hamiltonian,ZKM,Neven,Dumitrescu}. An analogous structure is found in Quantum Chromodynamics (QCD) for discretised fermions on a lattice~\cite{KieburgWilson,DSV,ADSV,MarioJacWilson}. Surprisingly in the latter example, the broadening of the zero eigenvalues coincides with the statistics of a finite-dimensional Gaussian random matrix model~\cite{KieburgWilson,DSV,ADSV,MarioJacWilson}, which have been corroborated by lattice simulations~\cite{DelDebbio:2005qa,DWW2011,DHS2012,CGRSZ}. These observations were surprising because universality of the spectral statistics, and thus agreement with RMT, usually only holds in the limit of a large number of eigenvalues, while the number of zero modes has been finite in these systems. A similar observation was found for outliers above the bulk of the spectrum, see, e.g., the mathematical review~\cite{Capitaine}. The statistics of outlier commonly play an enormous role in time series analysis and, thus, statistics~\cite{book}. In the present work, we want to investigate the mechanism behind these finite size universalities and we will see in Section \ref{Sec:Universality} that it is a mechanism similar to the Central Limit Theorem.

The main assumption needed to realise this is, in physical terms, that the zero modes are sufficiently delocalised in the eigenbasis of the perturbation.
We will consider average spectral properties, which could be an average over gauge fields, as in QCD, or an average over disorder in solid state systems.

In the present work, we model the physical ensemble average by an average over the Haar measure of the unitary matrix which expresses the unperturbed zero modes in the eigenbasis of the perturbation. This assumption is motivated by the fact that a perturbation that affects topology must be on a global scale. The short-distance dynamics of the corresponding modes are therefore averaged out.

It has been pointed out~\cite{BagretsAltland} that it is difficult to distinguish between accumulation of eigenvalues around the origin and perturbed topological modes in experiments. We propose to look at the different scaling behaviours of the eigenvalues and show that perturbed zero modes broaden with the system size in a way that is not shared by the bulk.
Our proposal is to exploit this difference as an indicator. The intuition behind this is that an accumulation of eigenvalues near zero will be part of the same ensemble as the first excited state, whereas perturbed zero modes behave as a separate finite-dimensional ensemble and therefore have a different scaling behaviour with the volume of the system and the coupling constant.
This scaling property was first observed for lattice QCD in~\cite{DelDebbio:2005qa} and understood within that context in~\cite{DSV,ADSV}. We show that it holds true for all ten symmetric spaces in the Altland-Zirnbauer classification and clarify the assumptions under which the $\nu\times\nu$ RMT behaviour of the near zero modes holds.

These results in the limit of large number of zero modes are also expected to be relevant for analysis of correlation matrices when applying a power map, see \cite{Powermap}.

Our starting point is a situation where a Hermitian operator $\hat{A}$ is perturbed by another Hermitian operator $\hat{S}$,
\begin{eqnarray}
\hat{K} = \hat{A} + \alpha \hat{S}. \label{Eq:ModelIntro}
\end{eqnarray}
We want to investigate the statistical properties of this operator, that is, the spectrum of eigenvalues upon an ensemble average. The coupling constant $\alpha$ will be chosen to be small such that first order perturbation theory can be applied. 
The procedure of the proof is as follows.

In Section \ref{Sec:Idea} we specify what is meant by ``small," where we also explain how to cut the Hilbert space to one of finite size $N$. The size $N$ will be sent to infinity at the end of the day.

We crystallise our assumptions in Section \ref{Sec:Model}, in particular the three conditions on $\hat{A}$, $\hat{S}$, and $\alpha$. For this purpose we show that the spectrum of the former zero modes decouple from the bulk for small $\alpha$. We also discuss that the first order perturbation theory becomes exact for $N\to\infty$ under the assumed conditions for all ten symmetry classes of Hermitian operators~\cite{Dyson,Martin,Casell,AlexMartin}.

In Section \ref{Sec:Universality} we then average over the part of the eigenbasis change between $\hat{A}$ and $\hat{S}$ associated with the zero modes of $\hat{A}$. The non-trivial change of basis creates a self average and forms a Central Limit Theorem for matrices. Our analysis deals with all ten symmetry classes in a unified way.

Our results are substantiated by numerics of some examples in Section \ref{Sec:Applications} that we expect will find some interest in lattice QCD and systems with Majorana modes in condensed matter system. In Section \ref{Sec:Conclusion} we conclude and discuss our results.

\section{Estimates of Scales}\label{Sec:Idea}

We start with a general unperturbed Hermitian operator $\hat{A}$. This operator might be a Hamiltonian, a Euclidean Dirac operator or another quantity. Due to its Hermiticity, we can decompose it in its eigenvalues $\lambda_j$ and its normalised eigenvectors $|\psi_j\rangle$, i.e.
\begin{equation}\label{specdec:A}
\hat{A}=\sum_j \lambda_j|\psi_j\rangle\langle\psi_j|.
\end{equation}
Here, we include degeneracies of the spectrum and zeros.
The operator may even have a continuum spectrum. In this case, we perform a finite volume UV cut-off for our analysis and let the volume $V$ go to infinity afterwards. Technically, we send the dimension $N$ of the Hilbert space to infinity, but the dimension is proportional to the volume of the system, $N\propto V$. This is true in QCD~\cite{KieburgWilson,DSV,ADSV,MarioJacWilson} and is expected to hold in condensed matter systems~\cite{AKMV,KimAdam} too. Usually, other quantities like the number of colours and the representation of the gauge group or the size of the spins and the number of particles enter into $N$ as well.

Let us assume that $\hat{A}$ has a fixed number $\nu>0$ of zero modes and the eigenvalues are ordered so that $|\lambda_k|>|\lambda_{N}|$ for all $k>N$ and $|\psi_j\rangle$ for $j=1,\dots,\nu$ form an orthonormal basis of the zero mode space. This ordering corresponds to the UV cut-off; the first $N$ eigenvalues are also the $N$ smallest. So we consider the truncated operator
\begin{equation}\label{trunc:A}
\hat{A}^{(N)}=\sum_{j=\nu+1}^N \lambda_j|\psi_j\rangle\langle\psi_j|.
\end{equation}
This operator may be represented by a matrix
\begin{equation}
\sum_j \lambda_j|\psi_j\rangle\langle\psi_j|\hat{=}\left(\begin{array}{c|c} A'=\diag(\lambda_{\nu+1},\ldots,\lambda_N) & 0 \\ \hline 0 & 0 \end{array}\right)
\end{equation}
The notation ``$\hat{=}$" will be used to indicate that the truncated operator in the eigenbasis of $\hat{A}$ is a finite-dimensional matrix.
We want to address how a generic additive Hermitian perturbation $\hat{S}$ broadens the eigenvalues of the zero modes for the operator
\begin{equation} \label{def:H}
\hat{K} = \hat{A} + \alpha \hat{S}=\lim_{N\to\infty} (\hat{A}^{(N)}+\alpha \hat{S}^{(N)})=\lim_{N\to\infty} \hat{K}^{(N)}
\end{equation}
with a small coupling constant $\alpha$ and the truncation of the perturbation $\hat{S}$ of the form
\begin{equation}\label{trunc:S}
\hat{S}^{(N)}=\sum_{j,k=1}^N\langle \psi_j|\hat{S}|\psi_k\rangle\ |\psi_j\rangle\langle \psi_k|.
\end{equation}
Note that $|\psi_j\rangle$ are still the eigenstates of $\hat{A}$.
Since we are only interested in the leading effects of $\hat{S}$ on the zero modes, we work in a perturbative regime. To this purpose, we  first need to identify what the correct scale of $\alpha$ is in terms of $\hat{A}$, $\hat{S}$, and $N$. Additionally, we have to specify how $\hat{S}$ describes a {\it generic perturbation}.

To get a feeling for the questions above, we do standard perturbation theory ignoring the fact that the spectra of $\hat{A}$ and $\hat{S}$ may vary over different scales. A more rigorous approach can be found in Section \ref{Sec:CondOp}.

The first order perturbation of the zero eigenvalues is given by the eigenvalues of the perturbation matrix
\begin{equation}\label{first_order}
\hat{K}_1^{\rm(zero)}=
\alpha \sum_{j',j=1}^\nu\langle \psi_j|\hat{S}|\psi_{j'}\rangle\ |\psi_j\rangle\langle \psi_{j'}|\ ,
\end{equation}
where the subscript denotes the order of the perturbation.
This perturbation is only dominant if it is smaller than the second order perturbation given by the eigenvalues of
\begin{equation}\label{second_order}
\hat{K}_2^{\rm(zero)}=-\alpha^2\sum_{j',j=1}^\nu\left(\sum_{k=\nu+1}^N\frac{\langle \psi_j|\hat{S}|\psi_k\rangle\langle \psi_k|\hat{S}|\psi_{j'}\rangle}{\lambda_k}\right)|\psi_j\rangle\langle \psi_{j'}|\ .
\end{equation}
The first and second order corrections are of equal magnitude when the largest singular value of $\hat{K}_2^{\rm(zero)}$ becomes of the same order as the smallest singular value of $\hat{K}_1^{\rm(zero)}$.
In Section \ref{Sec:Universality}, we argue that $\langle \psi_j|\hat{S}|\psi_{j'}\rangle$ are Gaussian distributed on the scale $\sqrt{\Tr  (\hat{S}^{(N)})^2}/N$ for large $N$ and sufficient mixing between the eigenbases of $\hat{A}$ and $\hat{S}$. The mixing is important for the Matrix Central Limit Theorem argument. The estimates of the smallest and largest singular value follow from, respectively,
\begin{equation}\label{heuristic:alpha.3}
\begin{split}
\left|\left|\left(\{\langle \psi_j|\hat{S}|\psi_{j'}\rangle\}_{i,j=1,\ldots,\nu}\right)^{-1}\right|\right|_{\rm op}\propto&\frac{\sqrt{\Tr  (\hat{S}^{(N)})^2}}{N},\\
\left|\left|\left\{\sum_{k=\nu+1}^N\frac{\langle \psi_j|\hat{S}|\psi_k\rangle\langle \psi_k|\hat{S}|\psi_{j'}\rangle}{\lambda_k}\right\}_{i,j=1,\ldots,\nu}\right|\right|_{\rm op}\leq& \frac{\Tr  (\hat{S}^{(N)})^2}{N^2|\lambda_{\nu+1}|}
\end{split}
\end{equation}
with $||.||_{\rm op}$ being the operator norm, meaning the largest singular value of the operator. From this we find the simple estimate
\begin{equation}\label{heuristic:alpha.4}
\frac{1}{N}\frac{\sqrt{\Tr  (\hat{S}^{(N)})^2}}{|\lambda_{\nu+1}|}\,\alpha\ll1
\end{equation}
for the coupling constant $\alpha$.
When the non-zero eigenvalues of $\hat{S}$ are of order $1$ and the smallest eigenvalue of $\hat{A}$ is of order $1/N$, we obtain $\sqrt{N}\alpha\ll1$, a relation which is well-known in lattice QCD~\cite{KieburgWilson,DSV,ADSV,MarioJacWilson}. Note that for certain ensembles the second order correction disappears due to symmetry. In this case we have to compare to the higher orders. This observation hints at the fact that we essentially need a different bound for $\alpha$ for the general situation. This is found in Section \ref{Sec:Model}. The discussion therein remains completely unaffected whether or not the second order perturbation theory vanishes.

As already mentioned, the heuristic approach above does not necessarily take into account that $\hat{A}$ as well as $\hat{S}$ may have several parts of their spectra that scale differently. Usually the smallest non-zero eigenvalue of $\hat{A}^{(N)}$ is of order $1/N$, see~\cite{VerbZahed,VerbaarschotThreeFold,JacBeta2,DOTV,LeutSmil,RMT_2_EFT-1,RMT_2_EFT-2,ADMNUniversality,Srednicki}. Moreover, the largest eigenvalue of $\hat{S}$ can even exceed the one of $\hat{A}$ as it is the case for the Wilson-Dirac operator~\cite{Wilson}. In such cases $\alpha$ can never be perturbative for the whole spectra but only for a certain subspectrum like the zero modes.
Equation~\eqref{heuristic:alpha.3} sets the scale where the perturbative approach of describing the broadening of the zero modes applies.

\section{Preparations}\label{Sec:Model}

The ensemble average we will consider is an average over the part of the transformation between the eigenbases of $\hat{A}$ and $\hat{S}$ associated with the zero modes. The full transformation is unitary and denoted by $U$. That is, diagonalising $\hat{S}^{(N)}=\sum_{l=1}^N s_l |\phi_l\rangle\langle\phi_l|$, we may write $U=\{\langle\psi_j|\phi_l\rangle\}_{j,l=1,\ldots,N}$.
The matrices $U$ will be drawn from the Haar measure of the group corresponding to the considered symmetry class, see Table~\ref{tab:quantities}. To motivate this form of the average, note that almost regardless what the eigenvalues $s_l$ are, the coefficients $\langle\psi_j|\phi_l\rangle\langle\phi_l|\psi_{j'}\rangle$ behave in a generic case like random variables. ``Generic" here means that these statements hold when averaging over the eigenvectors. We will later split $U$ into a part corresponding to the zero modes and a part corresponding to the rest of the spectrum.

Considering the leading order term $\hat{K}_1^{\rm(zero)}$ we note that each matrix entry can be expressed as a sum
\begin{equation}\label{Eq:Sum}
\langle\psi_j|\hat{S}|\psi_{j'}\rangle=\sum_{l=1}^N s_l \langle\psi_j|\phi_l\rangle\langle\phi_l|\psi_{j'}\rangle.
\end{equation}
The perturbation matrix for the zero modes is the part $j,j'=1,\ldots,\nu$. The Central Limit Theorem tells us that in the case of uncorrelated and identically distributed summands, the sum would be Gaussian. In Section \ref{Sec:Universality}, we extend the Central Limit Theorem to the sum \eqref{Eq:Sum} where neither the independence nor the identicalness is given. The fulcrum of our setup is that, for large $N$, the perturbation matrix for the zero modes becomes independent of the exact values of $s_l$. This requires the inverse participation ratio $\sum_{l=1}^{N}|\langle\psi_{j}|\phi_{l}\rangle|^4$ to be sufficiently small for $j=1,\dots,\nu$.  We show that all matrix entries with $j,j'=1,\ldots,\nu$ become Gaussian independent up to some symmetry relations due to this sum. That is, we show that this sum and, accordingly, the matrix entries are Gaussian. It hence follows that the eigenvalues obey a $\nu\times\nu$ Gaussian RMT.

We want to corroborate our statements from the previous section by listing the conditions under which the matrix valued Central Limit Theorem holds, see Subsection~\ref{Sec:CondOp}. Thereafter, in Subsection~\ref{Sec:Secular}, we explain why the first order perturbation theory becomes exact in the limit $N\to\infty$. Because the Central Limit Theorems depend on the symmetry class of the operators, we briefly review some of their particularities in Subsection~\ref{Sec:symmetryclass} and introduce our notation which is employed in Section \ref{Sec:Universality}.

\subsection{Conditions on the Operators}\label{Sec:CondOp}

We need the behaviour of the number of eigenvalues of $\hat{S}^{(N)}$ that are of the same order as its maximal singular value $\sigma_{\max}^{(N)}=||\hat{S}^{(N)}||_{\rm op}$ when $N$ goes to infinity. We recall that $||\hat{S}^{(N)}||_{\rm op}$ denotes the operator norm, meaning the largest singular value. A quantity which estimates the scaling of this number is the ratio between the $l^2$-norm and the operator norm,
\begin{equation}
q^{(N)}=\frac{\sqrt{\Tr  (\hat{S}^{(N)})^2}}{||\hat{S}^{(N)}||_{\rm op}}\in[1,\sqrt{N}].
\end{equation}
This quantity is akin to a participation ratio for eigenvalues.
With the help of this definition, we assume the following conditions
\begin{eqnarray}
\Tr  \hat{S}^{(N)} &=& 0, \label{Eq:CondTrace0}\\
\lim_{N\to\infty} q^{(N)} &=& \infty,\label{Eq:Gap}\\
\alpha &=& o\left(\frac{1}{||\hat{S}^{(N)}||_{\rm op}}\sqrt{\frac{N}{\Tr  (A')^{-2}}}\right).\label{Eq:strength}
\end{eqnarray}
The first condition is not mandatory but simplifies the notation below. If the trace does not vanish the whole spectrum is shifted by $\Tr  \hat{S}^{(N)}/N$. Hence, after a redefinition $\hat{S}^{(N)}-(\Tr  \hat{S}^{(N)}/N)\eins_N\to \hat{S}^{(N)}$ we end up with Equation \eqref{Eq:CondTrace0}. Additionally, it helps us avoid the completely degenerate case $\hat{S}\propto \eins$ where the Gaussian broadening of the zero modes collapses to a Dirac delta function (the spectrum is only shifted). This also shows that our results hold for any exact mode in a spectral gap.

The first true condition is Equation \eqref{Eq:Gap}. It guarantees the Gaussian random matrix approximation describing the broadening of the zero modes, see Section \ref{Sec:Universality}. Physically, the condition~\eqref{Eq:Gap} tells us that there are enough eigenvalues inducing self-averaging due to the the relative change of the eigenvectors of $\hat{A}$ and $\hat{S}$ for the Matrix Central Limit Theorem to apply. That is, there is sufficient delocalisation.
Note that this condition does not carry any information about the strength of the perturbation since the quotient $q^{(N)}$ is scale-invariant.

The bound on the strength of the perturbation is covered by condition~\eqref{Eq:strength}. It resembles Equation \eqref{heuristic:alpha.4} and describes when the first order approximation applies. One can show that Equation \eqref{Eq:strength} yields a stronger bound than Inequality \eqref{heuristic:alpha.4}, 
\begin{equation}\label{estimate}
\frac{N}{||(A')^{-1}||_{\rm op}\sqrt{\Tr  (\hat{S}^{(N)})^2}}\geq \frac{N}{q^{(N)}\sqrt{\Tr  (A')^{-2}}||\hat{S}^{(N)}||_{\rm op}}\geq \frac{\sqrt{N}}{\sqrt{\Tr  (A')^{-2}}||\hat{S}^{(N)}||_{\rm op}}.
\end{equation}
The stricter bound is necessary to truncate the perturbation series after the first term.
The interpretation is that $A'$ has to have a spectral gap where the former zero modes can live without being perturbed by the bulk.

\subsection{Secular Equation of the Broadened Zero Eigenvalues}\label{Sec:Secular}

Here we derive the first order perturbation from the secular equation of the whole system and study in detail the bounds for its validity.
As in Section \ref{Sec:Idea}, we choose to work in the eigenbasis of the truncated Hermitian operator $\hat{A}^{(N)}$. In this basis $\hat{S}^{(N)}$ takes the block form
(for the rest of our analysis, we represent the operators as $N\times N$ matrices $\hat{S}^{(N)}\hat{=}S^{(N)}$)
\begin{eqnarray}
US^{(N)}U\hc =\left(\begin{array}{c|c}
S_1 & S_2\\
\hline
S_2\hc & S_3
\end{array}\right).
\label{Eq:newbasis}
\end{eqnarray}
Here we have explicitly written the unitary matrix $U_{i,k}=\langle\psi_i|\phi_k\rangle$ which changes from the eigenbasis of  $S^{(N)}$ to $A^{(N)}$, that is
\begin{eqnarray}
[US^{(N)}U\hc]_{i,j} = U_{i,k}S^{(N)}_{k,k'}[U\hc]_{k',j} =\langle\psi_i|\phi_k\rangle\langle\phi_k|\Big(\sum_{l=1}^Ns_l|\phi_l\rangle\langle\phi_l|\Big)|\phi_{k'}\rangle\langle\phi_{k'}|\psi_j\rangle \ ,
\end{eqnarray}
where $k$ and $k'$ are summed over. Since the zero modes of $A^{(N)}$ make up the final $\nu$ rows of $U$ it is useful to introduce the symbol $U_2$ for this part of $U$, i.e., $[U]_{l,k}=[U_2]_{l,k}=\langle\psi_l|\phi_k\rangle$, where $l=N-\nu+1,\ldots,N$. Likewise we introduce the symbol $U_1$ for the first part of $U$, i.e., $[U]_{m,k}=[U_1]_{m,k}=\langle\psi_m|\phi_k\rangle$, where $m=1,\ldots,N-\nu$.

We do not make assumptions about the nature of these zero modes. They may be of topological origin, like anti-symmetry or chirality, or are given by peculiarities of the unperturbed system $\hat{A}$. Moreover, the symmetry classes of $\hat{A}$ and $\hat{S}$ are still open and will be discussed in the next subsection as well as in Section \ref{Sec:Universality}. Thence, we have not yet chosen the group $\mathcal{K}$ from where we draw the unitary matrix $U$ via the corresponding Haar measure, see Table~\ref{tab:quantities}.

To derive the first order perturbation of the secular equation of an eigenvalue $\lambda$, we start with the secular equation of the whole system $K^{(N)}=A^{(N)}+\alpha S^{(N)}$, i.e.
\begin{equation}
\det(K^{(N)}-\lambda\eins_N)=0.
\end{equation}
Employing the invariance of the determinant under the adjoint action of a unitary matrix we can rephrase this equation into the block form~\eqref{Eq:newbasis},
\begin{equation}\label{eq:sec.b}
\begin{split}
\det(K^{(N)}-\lambda \eins_N) =&\det\left(\begin{array}{c|c}
A'+\alpha S_1-\lambda\eins_{N-\nu} & \alpha S_2\\
\hline
\alpha S_2^\dagger & \alpha S_3-\lambda\eins_\nu
\end{array}\right)
\\
=& \det(A'-\lambda\eins_{N-\nu}) \det\left(\begin{array}{c|c}
\eins_{N-\nu} + \alpha (A'-\lambda\eins_{N-\nu})\inv S_1 & \alpha (A'-\lambda\eins_{N-\nu})\inv S_2\\
\hline
\alpha S_2^\dagger & \alpha S_3-\lambda\eins_\nu
\end{array}\right)
\\
=&\det\left(A' - \lambda\eins_{N-\nu}\right) \det\left(\eins_{N-\nu} + \alpha (A'-\lambda\eins_{N-\nu})\inv U_1 S U_1\hc \right)\\
&\times \det\left[\alpha U_2 S U_2\hc - \lambda\eins_\nu - \alpha U_2 S U_1\hc\left(\eins_{N-\nu} + \alpha\left(A' - \lambda \eins_{N-\nu}\right)\inv U_1 S U_1\hc \right)\inv \right.\\
&\times\left. \alpha \left(A' - \lambda\eins_{N-\nu}\right)\inv U_1 S U_2\hc \right]
\\
=&\det\left(A' - \lambda\eins_{N-\nu}\right)\det\left(\eins_N+\alpha S^{(N)}U_1^\dagger(A'-\lambda\eins_{N-\nu})\inv U_1\right)\\
&\times\det\left(\alpha U_2[\eins_N+\alpha S^{(N)}U_1^\dagger(A'-\lambda\eins_{N-\nu})\inv U_1]^{-1}S^{(N)}U_2^\dagger -\lambda \eins_\nu\right).
\end{split}
\end{equation}
In the second equality we pull out the factor $(A'-\lambda\eins_{N-\nu})$ in the first $N-\nu$ rows of the determinant. Then we have expanded the second determinant in its two blocks on the diagonal and exploited the explicit expression for $S_{1,2,3}$. The last line follows from the expression of inverse matrices as a Neumann series.

In the next step we make use of the bound of $\alpha$. Since the gap of $A'$ must not be allowed to close via the broadening of the zero modes, we need the smallest singular value of $A'$, which is $||(A')^{-1}||_{\rm op}$, to be much bigger than the largest singular value of $\alpha U_2[\eins_N+\alpha S^{(N)}U_1^\dagger(A')\inv U_1]^{-1}SU_2^\dagger$. Therefore, the dependence on $\lambda$ in the first two determinants of Equation \eqref{eq:sec.b} can be dropped so that those terms cannot vanish. This spectral gap between $A'$ and $\alpha U_2[\eins_N+\alpha S^{(N)}U_1^\dagger(A')\inv U_1]^{-1}S^{(N)}U_2^\dagger$ can most easily be seen when simplifying the latter. We can drop the term $\alpha S^{(N)}U_1^\dagger(A')\inv U_1$ because it is on average smaller than $\eins_N$. To see this let us choose an arbitrary vector $|\chi\rangle\in\mathbb{C}^N$. Then the square norm of $\alpha U_1^\dagger(A')\inv U_1S^{(N)}|\chi\rangle$ is on average
\begin{equation}
\int_{\mathcal{K}} d\mu(U)\alpha^2\langle\chi|S^{(N)}U_1^\dagger(A')^{-2} U_1 S^{(N)}|\chi\rangle=\frac{\alpha^2\Tr  (A')^{-2}}{N}\langle\chi|(S^{(N)})^2|\chi\rangle\leq\frac{\alpha^2\Tr  (A')^{-2}||S^{(N)}||_{\rm op}^2}{N}\ll1,
\end{equation}
where we used that each of the groups $\mathcal{K}$ comprises the symmetric group of permutations which immediately leads to the right hand side, cf.~Subsection~\ref{Sec:symmetryclass}. The second moment also vanishes as can be checked by
\begin{equation}
\begin{split}
\int_{\mathcal{K}} d\mu(U)\alpha^4(\langle\chi|S^{(N)}U_1^\dagger(A')^{-2} U_1 S^{(N)}|\chi\rangle)^2=&\frac{\alpha^4(c_1\Tr  (A')^{-4}+c_2(\Tr  (A')^{-2})^2)}{N^2}\langle\chi|(S^{(N)})^2|\chi\rangle^2\\
\leq&\frac{\alpha^4(|c_1|\Tr  (A')^{-4}+|c_2|(\Tr  (A')^{-2})^2)||S^{(N)}||_{\rm op}^4}{N^2},
\end{split}
\end{equation}
where $c_1$ and $c_2$ are two constants that are of order unity for large $N$.
Here, we used the fact that
\begin{equation}
\int_{\mathcal{K}} d\mu(U)|U_{ij}|^2|U_{il}|^2\overset{N\gg1}{\propto} \frac{1}{N^2},\ {\rm for}\ i,j,l=1,\ldots,N,
\end{equation}
for all of the groups $\mathcal{K}$ in Subsection~\ref{Sec:symmetryclass} and that $S^{(N)}|\chi\rangle\langle\chi|S^{(N)}$ is of rank one. Moreover we have $\Tr  (A')^{-4}\leq (\Tr  (A')^{-2})^2$ because $(A')^{-2}$ is positive definite. Hence, it holds
\begin{equation}
\int_{\mathcal{G}} d\mu(U)\alpha^4(\langle\chi|S^{(N)}U_1^\dagger(A')^{-2} U_1 S^{(N)}|\chi\rangle)^2\leq(|c_1|+|c_2|)\frac{\alpha^4(\Tr  (A')^{-2})^2||S^{(N)}||_{\rm op}^4}{N^2}\ll1.
\end{equation}
Therefore, on average each singular value of $\alpha S^{(N)}U_1^\dagger(A')\inv U_1$ is much smaller than unity and the term can be neglected in the sum $\eins_N+\alpha S^{(N)}U_1^\dagger(A')\inv U_1$.

Now we are ready to argue that $\lambda$ can be omitted in the combination $A'-\lambda\eins_{N-\nu}$ in the final determinant of \eqref{eq:sec.b}. This decouples the spectrum such that $\lambda$ measures the eigenvalues of
\begin{equation}\label{eq:secondmat}
\alpha U_2[\eins_N+\alpha S^{(N)}U_1^\dagger(A')\inv U_1]^{-1}S^{(N)}U_2^\dagger\approx\alpha U_2S^{(N)}U_2^\dagger.
\end{equation}
In Section \ref{Sec:Universality} we show that the matrix $U_2S^{(N)}U_2^\dagger$ is distributed according to a Gaussian random matrix where each matrix entry has the standard deviation $\sqrt{\Tr  (S^{(N)})^2}/N$. Due to the fixed and finite dimension $\nu$ (the number of the former zero modes), also the largest eigenvalue of the matrix \eqref{eq:secondmat} is of the order $\sqrt{\Tr  (S^{(N)})^2}/N$. We conclude that
\begin{equation}
\alpha \ll\frac{N}{||(A')^{-1}||_{\rm op}\sqrt{\Tr  (S^{(N)})^2}}
\end{equation}
is needed to drop $\lambda$ in $A'-\lambda\eins_{N-\nu}$. This is given from Equation \eqref{estimate}.

Summarising, with our assumed conditions~(\ref{Eq:CondTrace0}--\ref{Eq:strength}) the broadened zero modes are completely described by the leading order term $K_1^{\rm(z)}=\alpha S_3=\alpha U_2 S^{(N)}U_2^\dagger$. As we will show in Section \ref{Sec:Universality}, this matrix takes generically the form of a Gaussian random matrix.

\subsection{Symmetry Classes} \label{Sec:symmetryclass}

To see a broadening of finitely many zero modes we need an ensemble average. Otherwise we have only finitely many peaks somewhere about the origin. The ensemble average considered here will be an average over the matrix $U_2=\{\langle\psi_j|\phi_l\rangle\}_{j = N - \nu + 1,\dots,N, l = 1,\dots,N}$. We choose $U_2$ to be Haar-distributed in a Stiefel manifold of one of the groups $\mathcal{K}$ in Table \ref{tab:quantities}. Note that we do not require all of $U$ to be Haar-distributed.

The nature of the groups $\mathcal{K}$ strongly depends on what the generic symmetry class of $S_3=U_2SU_2^\dagger$ is. There are ten symmetry classes of Hermitian operators in total that $S_3$ can take. Those have been classified by Altland and Zirnbauer~\cite{Martin,AlexMartin}. Five of the ten classes exhibit a chiral symmetry and the other five do not. We start with the latter.

\subsubsection{Non-Chiral Classes}
The non-chiral symmetries can be described through the three number fields of real ($\mathbb{R}$), complex ($\mathbb{C}$), and quaternion ($\mathbb{H}$) numbers. These three fields each have a corresponding group, which are the orthogonal matrices ${\rm O}(N)$, the unitary matrices ${\rm U}(N)$, and the unitary symplectic matrices ${\rm USp}(N)$ with $N$ even. They are the maximal compact subgroups of the general linear groups $\mathcal{G}={\rm Gl}_{\mathbb{R}}(N),{\rm Gl}_{\mathbb{C}}(N),{\rm Gl}_{\mathbb{H}}(N)$, respectively. There are two Hermitian subsets invariant under ${\rm O}(N)$ which are the real symmetric matrices $\mathcal{H}={\rm Sym}(N)$ and the imaginary antisymmetric matrices  $\mathcal{H}={\rm ASym}(N)$. The same holds true for the quaternion case where we have the self-dual Hermitian matrices $\mathcal{H}={\rm Self}(N)$ and the anti-self-dual Hermitian matrices $\mathcal{H}={\rm ASelf}(N)$. For the complex case only the Hermitian matrices $\mathcal{H}={\rm Herm}(N)$ are invariant under ${\rm U}(N)$.  The matrix $S_3=U_2 S^{(N)}U_2^\dagger$ has to be in one of these five matrix sets when it is not generically chiral. Since only the projection of $U$ to its last $\nu$ rows is of interest, we do not average over the whole group $\mathcal{K}={\rm O}(N),{\rm U}(N),{\rm USp}(N)$ but only over the corresponding Stiefel manifolds $\mathcal{K}_\nu={\rm O}(N)/{\rm O}(N-\nu),{\rm U}(N)/{\rm U}(N-\nu),{\rm USp}(N)/{\rm USp}(N-\nu)$; for the last case also $\nu$ has to be even. In our calculations in Section \ref{Sec:Universality}, we need the fact that $\mathcal{K}_\nu$ can be embedded into $\nu\times N$ matrices which are given by the matrix spaces $\mathcal{G}_\nu={\rm Mat}_{\mathbb{R}}(\nu,N),{\rm Mat}_{\mathbb{C}}(\nu,N),{\rm Mat}_{\mathbb{H}}(\nu,N)$. We denote with $\mathcal{H}_\nu$ the matrix space from which $S_3$ is drawn.

\begin{table} 
\begin{tabular}{|c||c|c|c|c|}
  \hline
  RMT & Cartan Class & $\mathcal{H}_\nu$ & Matrix Structure \\
  \hline\hline
  GUE			
  & A & ${\rm Herm}(\nu)$ & $S_3=S_3^\dagger\in\mathbb{C}^{\nu\times \nu}$ 		\\ \hline
  GOE			
  & AI & ${\rm Sym}(\nu)$ & $\overset{\ }{S_3=S_3^T=S_3^*\in\mathbb{R}^{\nu\times \nu}}$		\\ \hline
  GSE			
  & AII & ${\rm Self}(\nu)$ & $\overset{\ }{\begin{array}{c} \displaystyle S_3=\tau_2S_3^T\tau_2=\tau_2S_3^*\tau_2\in\mathbb{C}^{\nu\times \nu},\ \nu\in2\mathbb{N}\end{array}}$		\\ \hline
  GAOE			
  & B$\mid$D & ${\rm ASym}(\nu)$ & $\overset{\ }{\begin{array}{c} \displaystyle S_3=-S_3^T=-S_3^*\in\imath\mathbb{R}^{\nu\times\nu}\end{array}}$		\\ \hline
  GASE			
  & C & ${\rm ASelf}(\nu)$ & $\overset{\ }{\begin{array}{c} \displaystyle S_3=-\tau_2S_3^T\tau_2=-\tau_2S_3^*\tau_2\in\mathbb{C}^{\nu\times\nu},\ \nu\in2\mathbb{N}\end{array}}$		\\ \hline\hline
  $\chi$GUE			
  & AIII & ${\rm Mat}_{\mathbb{C}}(p',n')$ & $\overset{\ }{\begin{array}{c} \displaystyle S_3=\left[\begin{array}{cc} 0 & W_3 \\ W_3^\dagger & 0 \end{array}\right],\ W_3\in\mathbb{C}^{p'\times n'}\end{array}}$	\\ \hline
  $\chi$GOE			
  & B$\mid$DI & ${\rm Mat}_{\mathbb{R}}(p',n')$ & $\overset{\ }{\begin{array}{c} \displaystyle S_3=\left[\begin{array}{cc} 0 & W_3 \\ W_3^\dagger & 0 \end{array}\right],\ W_3=W_3^*\in\mathbb{R}^{p'\times n'}\end{array}}$	\\ \hline
  $\chi$GSE			
  & CII & ${\rm Mat}_{\mathbb{H}}(p',n')$ &  $\overset{\ }{\begin{array}{c} \displaystyle S_3=\left[\begin{array}{cc} 0 & W_3 \\ W_3^\dagger & 0 \end{array}\right],\ W_3=\tau_2W^*\tau_2\in\mathbb{C}^{p'\times n'},\ p',n'\in2\mathbb{N}\end{array}}$		\\ \hline\hline
  GBOE			
  & CI & ${\rm Sym}_{\mathbb{C}}(\nu/2)$ & $\overset{\ }{\begin{array}{c} \displaystyle S_3=\left[\begin{array}{cc} 0 & W_3 \\ W_3^\dagger & 0 \end{array}\right],\ W_3=W_3^T\in\mathbb{C}^{\nu/2\times \nu/2},\ \nu\in2\mathbb{N} \end{array}}$		\\ \hline
  GBSE			
  & DIII & ${\rm ASym}_{\mathbb{C}}(\nu/2)$ & $\overset{\ }{\begin{array}{c} \displaystyle S_3=\left[\begin{array}{cc} 0 & W_3 \\ W_3^\dagger & 0 \end{array}\right],\ W_3=-W_3^T\in\mathbb{C}^{\nu/2\times \nu/2},\ \nu\in2\mathbb{N} \end{array}}$		\\ \hline
\end{tabular}
\caption{The ten symmetry classes given in terms of the acronym of the Gaussian random matrix ensemble (first column, notation follows~\cite{MarioTim}) and the symbol along the Cartan classification scheme (second column, see~\cite{Martin,Casell,AlexMartin}). The third column represents the matrix space in which $S_3$ lives, and the fourth column shows its structure in matrix form. The matrix $\tau_2$ is the second Pauli matrix. In the first five rows we listed the non-chiral classes followed by the three classical chiral ensembles where $p+n=N$ and $p'+n'=\nu$. The two Boguliubov--de Gennes classes are given in the last two rows. For the symplectic cases (third, fifth and eighth row) the dimensions $N,\nu, p,\ldots$ have to be all even. This table is continued in Table~\ref{tab:quantities}.
}\label{tab:symmetryclasses}
\end{table}

\begin{table} 
\begin{tabular}{|c||c|c|c|c|}
  \hline
  RMT & $\mathcal{K}_\nu$ & $\mathcal{G}_\nu$ & $\mathcal{P}_\nu$ & $\gamma$ \\
  \hline\hline
  GUE			
  & $\displaystyle\frac{{\rm U}(N)}{{\rm U}(N-\nu)}$ & ${\rm Mat}_{\mathbb{C}}(\nu,N)$ &  ${\rm Herm}(\nu)$ & $1$ 		\\ \hline
  GOE			
  & $\displaystyle\frac{{\rm O}(N)}{{\rm O}(N-\nu)}$ & ${\rm Mat}_{\mathbb{R}}(\nu,N)$ &  ${\rm Sym}(\nu)$ & $1/2$		\\ \hline
  GSE			
  & $\displaystyle\frac{{\rm USp}(N)}{{\rm USp}(N-\nu)}$ & ${\rm Mat}_{\mathbb{H}}(\nu,N)$ &  ${\rm Self}(\nu)$ & $1/2$		\\ \hline
  GAOE			
  & $\displaystyle\frac{{\rm O}(N)}{{\rm O}(N-\nu)}$ & ${\rm Mat}_{\mathbb{R}}(\nu,N)$ &  ${\rm Sym}(\nu)$ & $1/2$		\\ \hline
  GASE			
  & $\displaystyle\frac{{\rm USp}(N)}{{\rm USp}(N-\nu)}$ & ${\rm Mat}_{\mathbb{H}}(\nu,N)$ & ${\rm Self}(\nu)$ & $1/2$		\\ \hline\hline
  $\chi$GUE			
  & $\displaystyle\frac{{\rm U}(p)}{{\rm U}(p-p')}\times\frac{{\rm U}(n)}{{\rm U}(n-n')}$ &  ${\rm Mat}_{\mathbb{C}}(p',p)\oplus{\rm Mat}_{\mathbb{C}}(n',n)$ &  ${\rm Herm}(p')\oplus{\rm Herm}(n')$ & $1$	 \\ \hline
  $\chi$GOE			
  & $\displaystyle\frac{{\rm O}(p)}{{\rm O}(p-p')}\times \frac{{\rm O}(n)}{{\rm O}(n-n')}$ & ${\rm Mat}_{\mathbb{R}}(p',p)\oplus{\rm Mat}_{\mathbb{R}}(n',n)$ & ${\rm Sym}(p')\oplus{\rm Sym}(n')$ & $1/2$	\\ \hline
  $\chi$GSE			
  & $\displaystyle\frac{{\rm USp}(p)}{{\rm USp}(p-p')}\times\frac{{\rm USp}(n)}{{\rm USp}(n-n')}$ & ${\rm Mat}_{\mathbb{H}}(p',p)\oplus{\rm Mat}_{\mathbb{H}}(n',n)$ & ${\rm Self}(p')\oplus{\rm Self}(n')$ & $1/2$		\\ \hline\hline
  GBOE			
  & $\displaystyle\frac{{\rm U}(N/2)}{{\rm U}((N-\nu)/2)}$ & ${\rm Mat}_{\mathbb{C}}(\nu/2,N/2)$ & ${\rm Herm}(\nu/2)$ & $1/2$		\\ \hline
  GBSE			
  & $\displaystyle\frac{{\rm U}(N/2)}{{\rm U}((N-\nu)/2)}$ &${\rm Mat}_{\mathbb{C}}(\nu/2,N/2)$ & ${\rm Herm}(\nu/2)$ & $1/2$		\\ \hline
\end{tabular}
\caption{ Continuation of Table~\ref{tab:symmetryclasses} where the order of the rows is the same.  The first column shows again the acronym of the corresponding ensemble. The corresponding Stiefel manifold, which we obtain after projecting the eigenvectors   $U=\{\langle\psi_j|\phi_l\rangle\}_{j,l=1,\ldots,N}$ to the broadened zero modes $U_2$, is given in the second column, and the third column shows the flat matrix space in which $\mathcal{K}_\nu$ is embedded. This embedding is needed in our calculations in Section \ref{Sec:Universality}. The same is also true for the Hermitian matrix spaces given in the fourth column, that are employed to rewrite the Haar measures as Gaussian integrals. The parameter $\gamma$ in the last column appears at several places in the derivation too. It is essentially the exponent of the determinant that can be obtained by a multivariate Gaussian integral.
}\label{tab:quantities}
\end{table}

\subsubsection{Chiral Classes}
When chiral symmetry is present the situation is slightly more complicated. There are the three standard chiral symmetry classes~\cite{VerbaarschotThreeFold}, where
\begin{equation}\label{chiral}
\hat{S}^{(N)}\hat{=}\left(\begin{array}{cc} 0 & W \\ W^\dagger & 0 \end{array}\right)
\end{equation}
comprises a real ($W\in {\rm Mat}_{\mathbb{R}}(p,n)\hat{=}\mathcal{H}$), complex  ($W\in {\rm Mat}_{\mathbb{C}}(p,n)\hat{=}\mathcal{H}$), or a quaternion  ($W\in {\rm Mat}_{\mathbb{H}}(p,n)\hat{=}\mathcal{H}$ with $p$ and $n$ even) matrix with $p+n=N$. Here the notion ``$\hat{=}$" carries the additional meaning that there is a unitary matrix for the ensemble where $S^{(N)}$ is drawn from to write it in this form. The matrix $U=\diag(V_1,V_2)$ can be chosen in a block diagonal form with $(V_1,V_2)\in\mathcal{K}={\rm O}(p)\times{\rm O}(n),{\rm U}(p)\times{\rm U}(n),{\rm USp}(p)\times{\rm USp}(n)$.

The remaining two symmetry classes are of the Bogoliubov--de Gennes type where $W$ is either complex symmetric ${\rm Sym}_{\mathbb{C}}(p=n=N/2)\hat{=}\mathcal{H}$ or complex antisymmetric ${\rm ASym}_{\mathbb{C}}(p=n=N/2)\hat{=}\mathcal{H}$. In both cases the unitary group $\mathcal{K}={\rm U}(N/2)$ keeps this structure invariant, but the unitary matrix $U=\diag(V_1,V_2)$ satisfies the condition $V_1=V_2^*$.

To get the statistics of the cut-out $S_3$ we assume that the projection is symmetry-preserving, meaning $S$ and $S_3$ share the same symmetry class though they are of different dimensions. The matrix $S_3$ should be also chiral,
\begin{equation}\label{chiral.b}
S_3\hat{=}\left(\begin{array}{cc} 0 & W_3 \\ W_3^\dagger & 0 \end{array}\right)
\end{equation}
with $W_3$ being $p'\times n'$ dimensional, where the dimensions satisfy $p'\leq p$, $n'\leq n$, and $p'+n'=\nu\leq N$. Due to this projection we have to effectively integrate over the Stiefel manifolds $\mathcal{K}_\nu={\rm O}(p)/{\rm O}(p-p')\times {\rm O}(n)/{\rm O}(n-n'),{\rm U}(p)/{\rm U}(p-p')\times{\rm U}(n)/{\rm U}(n-n'),{\rm USp}(p)/{\rm USp}(p-p')\times{\rm USp}(n)/{\rm USp}(n-n')$ for the three classical chiral ensembles. As for the non-chiral ensembles we need their embedding in a flat vector space which here is $\mathcal{G}_\nu={\rm Mat}_{\mathbb{R}}(p',p)\oplus{\rm Mat}_{\mathbb{R}}(n',n),{\rm Mat}_{\mathbb{C}}(p',p)\oplus{\rm Mat}_{\mathbb{C}}(n',n),{\rm Mat}_{\mathbb{H}}(p',p)\oplus{\rm Mat}_{\mathbb{H}}(n',n)$. For the two Boguliubov--de Gennes classes the two spaces are $\mathcal{K}_\nu={\rm U}(N/2)/{\rm U}((N-\nu)/2)$ and $\mathcal{G}_\nu={\rm Mat}_{\mathbb{C}}((N-\nu)/2,N/2)$. Here let us emphasise that for these two cases $N$ as well as $\nu$ are assumed to be even.

The above discussion is summarised in Tables~\ref{tab:symmetryclasses} and~\ref{tab:quantities}.

\section{Central Limit Theorems for Matrices}\label{Sec:Universality}

In this section, we want to answer the question  what the distribution of the matrix $S_3 = U_2S^{(N)}U_2^\dagger$ of finite size $\nu\times \nu$ is when $N$ becomes large. We here ignore the overall factor $\alpha$ as the perturbative expansion of the zero modes has already taken place, see Subsection~\ref{Sec:Secular}. We study the non-chiral, the classical chiral, and the Bogoliubov--de Gennes classes separately in Subsections~\ref{Sec:CondUn1},~\ref{Sec:CondUn2}, and~\ref{Sec:CondUn2}. For all ten symmetry classes we find  that under the conditions~\CondS\ $S_3$ is distributed by a Gaussian  in the limit of large $N$.
Results from effective field theory~\cite{MarioJacWilson,KimAdam} suggest that these results hold for an even more general setting when the unitary submatrix $U_2$ is not Haar distributed.

\subsection{Gaussian Limit for Non-Chiral $S_3$} \label{Sec:CondUn1}

We define the distribution of $S'=\kappa S_3$, with $\kappa=N/\sqrt{\Tr  (S^{(N)})^2}$, via a Dirac delta function,
\begin{equation}\label{Eq:DensSbar}
p(S') = \int_{\mathcal{K}_\nu} d\mu(U_2) \delta\left(S' - \kappa U_2S^{(N)}U_2^\dagger\right),
\end{equation}
where $d\mu(U_2)$ is the normalised Haar measure of the Stiefel manifold $\mathcal{K}_\nu$, see the first five rows of Tables~\ref{tab:symmetryclasses} and~\ref{tab:quantities}. We have contained the scaling explicitly in $\kappa$ to simplify later calculations.
The Haar measure has also a representation as a Dirac delta function over the larger set $\mathcal{G}_\nu$,
\begin{equation}\label{Eq:Haar}
 \int_{\mathcal{K}_\nu}d\mu(U_2) f(U_2) =\frac{\int_{\mathcal{G}_\nu} dU_2 f(U_2)\delta(\eins_\nu-U_2U_2^\dagger)}{\int_{\mathcal{G}_\nu} dU_2 \delta(\eins_\nu-U_2U_2^\dagger)}.
\end{equation}
with an arbitrary integrable function $f$.
Both Dirac delta functions can be expressed as Gaussian integrals over the symmetric spaces $\mathcal{H}_\nu$ for Equation \eqref{Eq:DensSbar} and $\mathcal{P}_\nu$ for Equation \eqref{Eq:Haar}. Thus, we start with the expression
\begin{equation}\label{calc:nonchi.1}
\begin{split}
p(S')=&\lim_{\epsilon\to0} \frac{\int_{\mathcal{G}_\nu} dU_2\int_{\mathcal{P}_\nu}dP f_\epsilon(U_2,S')\exp[ \epsilon \gamma N\Tr  (\eins_\nu-i P)^2+\gamma N\Tr (\eins_\nu-U_2U_2^\dagger)(\eins_\nu-i P)]}{\int_{\mathcal{G}_\nu} dU_2\int_{\mathcal{P}_\nu}dP\exp[ \epsilon \gamma N\Tr  (\eins_\nu-i P)^2+\gamma N\Tr (\eins_\nu-U_2U_2^\dagger)(\eins_\nu-i P)]},\\
f_\epsilon(U_2,S')=&\frac{\int_{\mathcal{H}_\nu}dH\exp[ -\epsilon\Tr  H^2+i\Tr (S' - \kappa U_2S^{(N)}U_2^\dagger)H]}{\int_{\mathcal{H}_\nu}d\bar{S}\int_{\mathcal{H}_\nu}dH\exp[- \Tr  H^2- \Tr  \bar{S}^2/4]}
\end{split}
\end{equation}
to analyse the large $N$ behaviour. The shift in $H$ guarantees that the integral over $U_2$ is absolutely integrable and the denominators normalize the integrals properly. The factor $\gamma N$ in the $P$-dependent part of the exponent is introduced in foresight of the saddle point approximation when taking $N\to\infty$. Here $\gamma $ is a parameter depending on the symmetry class and can be read off from Table~\ref{tab:quantities}.

Due to the absolute integrability of the integrals we can interchange them. This allows us to carry out the integral over $U_2$ which is now a Gaussian over a $\nu\times N$ dimensional matrix yielding a determinant. Thence, we find
\begin{equation}\label{calc:nonchi.2}
\begin{split}
p(S')=&\lim_{\epsilon\to0} \frac{\int_{\mathcal{P}_\nu}dP\widetilde{f}_\epsilon(P,S')\exp[\epsilon \gamma N\Tr  (\eins_\nu-i P)^2+\gamma N\Tr (\eins_\nu-i P)]}{\int_{\mathcal{P}_\nu}dP\exp[\epsilon \gamma N\Tr  (\eins_\nu-i P)^2+\gamma N\Tr (\eins_\nu-i P)]\det^{-\gamma N}[\gamma N (\eins_\nu-iP)]},\\
\widetilde{f}_\epsilon(P,S')=&\frac{\int_{\mathcal{H}_\nu}dH\exp[-\epsilon\Tr  H^2+i\Tr  S'H]\det^{-\gamma}[\gamma N\eins_N\otimes (\eins_\nu-iP)+i \kappa S^{(N)}\otimes H]}{\int_{\mathcal{H}_\nu}d\bar{S}\int_{\mathcal{H}_\nu}dH\exp[- \Tr  H^2- \Tr  \bar{S}^2/4]},
\end{split}
\end{equation}
where the exponent $\gamma$ depends on the symmetry class and can be read off from Table~\ref{tab:quantities}.

For $N$ large enough, the limit $\epsilon\to0$ can be performed for the integral over $P$ because the determinant guarantees the convergence. However, we still need this regularisation for the integral over $H$. We therefore do the saddle point analysis of the simplified version
\begin{equation}\label{calc:nonchi.3}
\begin{split}
p(S')=& \frac{\int_{\mathcal{P}_\nu}dP g(P,S')\exp[ -i\gamma N\Tr  P]\det^{-\gamma N}[\eins_\nu-iP]}{\int_{\mathcal{P}_\nu}dP\exp[-i\gamma N\Tr  P]\det^{-\gamma N}[\eins_\nu-iP]},\\
g(P,S')=&\lim_{\epsilon\to0}\frac{\int_{\mathcal{H}_\nu}dH\exp[-\epsilon\Tr  H^2+i\Tr  S'H]\det^{-\gamma}[\eins_{N\nu} +i\gamma^{-1} S^{(N)}/\sqrt{\Tr (S^{(N)})^2}\otimes H(\eins_\nu-iP)^{-1}]}{\int_{\mathcal{H}_\nu}d\bar{S}\int_{\mathcal{H}_\nu}dH\exp[- \Tr  H^2- \Tr  \bar{S}^2/4]},
\end{split}
\end{equation}
where we have written out $\kappa$. For large $N$, we rescale $P\to P/\sqrt{\gamma N}$ in the enumerator as well as in the denominator. This allows us to perform the limit for the $P$ integral exactly with Lebesgue's dominated convergence theorem. We have also written out $\kappa$. This implies that the $P$-integrand becomes the Gaussian $\exp[-\Tr  P^2/2]$ via a Taylor expansion. Hence we obtain
\begin{equation}\label{calc:nonchi.4}
\begin{split}
\lim_{N\to\infty}p(S')=& \lim_{N\to\infty}\lim_{\epsilon\to0}\frac{\int_{\mathcal{H}_\nu}dH\exp[-\epsilon\Tr  H^2+i\Tr  S'H]\det^{-\gamma}[\eins_{N\nu} +i\gamma^{-1} S^{(N)}/\sqrt{\Tr (S^{(N)})^2}\otimes H]}{\int_{\mathcal{H}_\nu}d\bar{S}\int_{\mathcal{H}_\nu}dH\exp[- \Tr  H^2- \Tr  \bar{S}^2/4]}.
\end{split}
\end{equation}

The limit of the integral over $H$ results from an expansion of the determinant which is
\begin{equation}\label{calc:nonchi.5}
{\rm ln}\,{\det}^{-\gamma}\left[\eins_{N\nu} +i \frac{S^{(N)}}{\gamma\sqrt{\Tr (S^{(N)})^2}}\otimes H\right]=\gamma\sum_{j=1}^\infty\frac{1}{j}\Tr \left(-i \frac{S^{(N)}}{\gamma\sqrt{\Tr (S^{(N)})^2}}\right)^j \Tr  H^j.
\end{equation}
The first term ($j=1$) vanishes because of condition~\eqref{Eq:CondTrace0} and the coefficient for $j=2$ becomes $-1/(2\gamma)$. The other terms for $j>2$ can be estimated as follows,
\begin{equation}\label{calc:nonchi.6}
\left|\frac{\Tr(S^{(N)})^j}{(\Tr (S^{(N)})^2)^{j/2}}\right|\leq\frac{||S^{(N)}||_{\rm op}^{j-2}\Tr (S^{(N)})^2}{(\Tr (S^{(N)})^2)^{j/2}}=\frac{1}{(q^{(N)})^{j-2}}\overset{N\to\infty}{\rightarrow}0
\end{equation}
resulting from the condition \eqref{Eq:Gap}. Therefore, the determinant can be approximated by a Gaussian telling us that we can set $\epsilon=0$. Eventually we arrive at
\begin{equation}\label{result:non-chiral}
\begin{split}
\lim_{N\to\infty}p(S')=& \frac{\int_{\mathcal{H}_\nu}dH\exp[-\Tr  H^2/(2\gamma)+i\Tr  S'H]}{\int_{\mathcal{H}_\nu}d\bar{S}\int_{\mathcal{H}_\nu}dH\exp[- \Tr  H^2- \Tr  \bar{S}^2/4]}=\frac{\exp[-\gamma\Tr  {S'}^2/2]}{\int_{\mathcal{H}_\nu}d\bar{S}\exp[- \gamma\Tr  \bar{S}^2/2]},
\end{split}
\end{equation}
which is the main result of the section.

We conclude that the former zero eigenvalues are broadened by the matrix $\alpha S_3$ which is distributed like a Gaussian random matrix with standard deviation $\alpha\sqrt{\Tr  (S^{(N)})^2/(\gamma N^2)}$ for large $N$. 

\subsection{Gaussian Limit of  $S_3$ for one of the three Standard Chiral Classes} \label{Sec:CondUn2}

The three classical chiral ensembles can be dealt with in a similar way to the five non-chiral ensembles in the previous section. We anew replace the normalised Haar measure of $\mathcal{K}_\nu$ by a Gaussian integral over $\mathcal{G}_\nu$ and $\mathcal{P}_\nu$ and the Dirac delta function in $S'$ by a Gaussian integral on $\mathcal{H}_\nu$. Thus, Equation~\eqref{calc:nonchi.1} still holds only for the respective spaces, see the sixth to eighth row of the Tables~\ref{tab:symmetryclasses} and~\ref{tab:quantities}. The difference shows in the structure of the matrices. While the matrix $\gamma N (\eins_N-iP)=\diag(\gamma p(\eins_{p'}-iP_1),\gamma n (\eins_{n'}-iP_2))$ is block diagonal, one block is of size $p'\times p'$ and the other of size $n'\times n'$, the matrices
\begin{equation}\label{calc:chi.1}
S^{(N)}=\left(\begin{array}{cc} 0 & W \\ W^\dagger & 0 \end{array}\right)\ {\rm as\ well\ as}\ H=\left(\begin{array}{cc} 0 & X \\ X^\dagger & 0 \end{array}\right)
\end{equation}
consist of off-diagonal blocks of size $p\times n$ and $n\times p$ as well as $p'\times n'$ and $n'\times p'$, respectively. Note that we weight the two blocks of $P$ differently, again in foresight of the saddle point analysis. With this in mind one can perform the integral over $U_2=\diag(\widetilde{V}_1,\widetilde{V}_2)$ leading to the counterpart of Equation \eqref{calc:nonchi.2} with the appropriate matrix spaces and the exponent $\gamma$ as given in Table~\ref{tab:quantities}. Here we use the identity
\begin{equation}\label{calc:chi.2}
\begin{split}
&\int_{\mathcal{G}_\nu}d(\widetilde{V}_1,\widetilde{V}_2)\exp\Big[-\gamma p\Tr \widetilde{V}_1^\dagger(\eins_{p'}-iP_1) \widetilde{V}_1-\gamma n\Tr \widetilde{V}_2^\dagger (\eins_{n'}-iP_2) \widetilde{V}_2\\
&-i\kappa\Tr \widetilde{V}_1^\dagger X \widetilde{V}_2 W^\dagger-i\kappa\Tr\widetilde{V}_2^\dagger X^\dagger \widetilde{V}_1 W\Big]\\
\propto&\ {\det}^{-\gamma}\left[\begin{array}{cc} \gamma p\eins_p\otimes (\eins_{p'}-iP_1) & i\kappa W\otimes X \\ i\kappa W^\dagger\otimes X^\dagger & \gamma n\eins_n\otimes (\eins_{n'}-iP_2) \end{array}\right],
\end{split}
\end{equation}
which can be readily computed.

The rest of the calculation does not differ much from the non-chiral situation. First we can take the limit $\epsilon\to0$ in the $P$-integral because the convergence is given by the determinant and the limit $N\to\infty$, which implies that $p/N$ and $n/N$ are fixed since the number of zero modes shall be fixed, can be done for $P$ exactly after rescaling $P_1\to P_1/\sqrt{\gamma p}$ and  $P_2\to P_2/\sqrt{\gamma n}$. Finally, we expand the remaining determinant,
\begin{equation}\label{calc:chi.3}
\begin{split}
&{\det}^{-\gamma}\left[\eins_{N\nu}+\left(\begin{array}{cc} 0 & \displaystyle i\frac{\kappa}{\gamma\sqrt{pn}} W\otimes X \\ \displaystyle i\frac{\kappa}{\gamma\sqrt{pn}} W^\dagger\otimes X^\dagger & 0 \end{array}\right)\right]\\
=&\gamma\sum_{j=1}^\infty\frac{1}{j}\Tr\left(-\frac{N^2}{\gamma^2 pn \Tr(S^{(N)})^2} WW^\dagger\right)^j\Tr(XX^\dagger)^j.
\end{split}
\end{equation}
In view of $2\Tr (WW^\dagger)^j=\Tr(S^{(N)})^{2j}$, we can exploit the same estimation as in Equation \eqref{calc:nonchi.6} such that only the term for $j=1$ survives. The leftover Gaussian integral over $H$ can be carried out and we obtain the result
\begin{equation}\label{result:chiral}
\begin{split}
\lim_{N\to\infty}p(S')=& \frac{\exp[-\gamma pn\Tr  {S'}^2/N^2]}{\int_{\mathcal{H}_\nu}d\bar{S}\exp[- \gamma pn\Tr  \bar{S}^2/N^2]}.
\end{split}
\end{equation}
Consequently, the matrix $\alpha S_3$ is again distributed along a Gaussian random matrix with a standard deviation $\alpha\sqrt{\Tr(S^{(N)})^2/(2\gamma pn)}$.

\subsection{Gaussian Limit for the Boguliubov--de Gennes types of $S_3$} \label{Sec:CondUn3}

For the two Boguliubov--de Gennes cases we have almost the same situation as in other three chiral classes only that for $U_2=\diag(\widetilde{V}_1,\widetilde{V}_2)$ we have additionally the condition $\widetilde{V}_2=\widetilde{V}_1^*$. For this reason, the matrix $P$ satisfies the diagonal block form $P=\diag(\widetilde{P},\widetilde{P}^*)$ with $\widetilde{P}\in{\rm Herm}(\nu/2)$. The matrices $S^{(N)}$ and $H$ attain the chiral forms~\eqref{calc:chi.1} with the additional conditions $W^T=\pm W$ and $X^T=\pm X$, both relations with the same sign.

Starting with Equation \eqref{calc:nonchi.1} only with the corresponding matrix spaces, see last two rows of the Tables~\ref{tab:symmetryclasses} and~\ref{tab:quantities}, as well as replacing $N$ by $N/2$ and setting $\gamma=1/2$ in the exponential functions, we need the counterpart of Equation \eqref{calc:chi.2} which is
\begin{equation}\label{calc:BdG.1}
\begin{split}
&\int_{\mathcal{G}_\nu}d\widetilde{V}_1\exp\left[-N\Tr \widetilde{V}_1^\dagger(\eins_{\nu/2}-i\widetilde{P}) \widetilde{V}_1/2-i\kappa\Tr \widetilde{V}_1^\dagger X \widetilde{V}_1^* W^\dagger-i\kappa\Tr \widetilde{V}_1^T X^\dagger \widetilde{V}_1 W\right]\\
\propto&\ {\det}^{-1/2}\left[\frac{N}{2}\eins_2\otimes\eins_{N/2}\otimes (\eins_{\nu/2}-i\widetilde{P})+i\kappa(\tau_3-i\tau_1)\otimes X\otimes W^\dagger+i\kappa(\tau_3+i\tau_1)\otimes X^\dagger\otimes W\right]\\
=&\ {\det}^{-1/2}\left[\begin{array}{cc} \displaystyle\frac{N}{2}\eins_{N/2}\otimes (\eins_{\nu/2}-i\widetilde{P}) & 2i\kappa X^\dagger \otimes W \\ 2i\kappa X\otimes W^\dagger & \displaystyle\frac{N}{2}\eins_{N/2}\otimes (\eins_{\nu/2}-i\widetilde{P}) \end{array}\right],
\end{split}
\end{equation}
with $\tau_j$ the three Pauli matrices. The second line is obtained after decomposing $\widetilde{V}_1$ into real and imaginary part and the third line can be found by performing a rotation with $\exp[i\pi(\eins_2-\tau_3)/4]\exp[i \pi \tau_2/4]$. The saddle point expansion can be achieved by rescaling $\widetilde{P}\to \widetilde{P}/\sqrt{\gamma N}$ and the Taylor expansion of the determinant works along Equation \eqref{calc:chi.3}. We hereby again find the Gaussian distribution
\begin{equation}\label{result:BdG}
\begin{split}
\lim_{N\to\infty}p(S')=& \frac{\exp[-\Tr  {S'}^2]}{\int_{\mathcal{H}_\nu}d\bar{S}\exp[- \Tr  \bar{S}^2]},
\end{split}
\end{equation}
which implies that $\alpha S_3$ is a Gaussian random matrix with standard deviation $\alpha\sqrt{\Tr(S^{(N)})^2/(2N^2)}$ in the limit $N\to\infty$.

\section{Scaling and Application}\label{Sec:Applications}
Let us analyse the scaling behaviour of the spectra in more detail. As mentioned above, the smallest eigenvalue of $A^{(N)}$ is typically on the scale $N\inv$. We may therefore zoom in on the microscopic spectrum around the origin if we consider rescaled eigenvalues
\begin{eqnarray}
x = N \lambda\ ,\label{Eq:EigRescale}
\end{eqnarray}
where $\lambda$ are the eigenvalues of $K^{(N)}$, in the limit $N\to \infty$.

\begin{figure}
	\hspace{-0.39\linewidth}	\textbf{(a)}\hspace{0.47\linewidth}	\textbf{(b)}\hfill\\
	\centering
		\includegraphics[width=0.49\linewidth,angle=0]{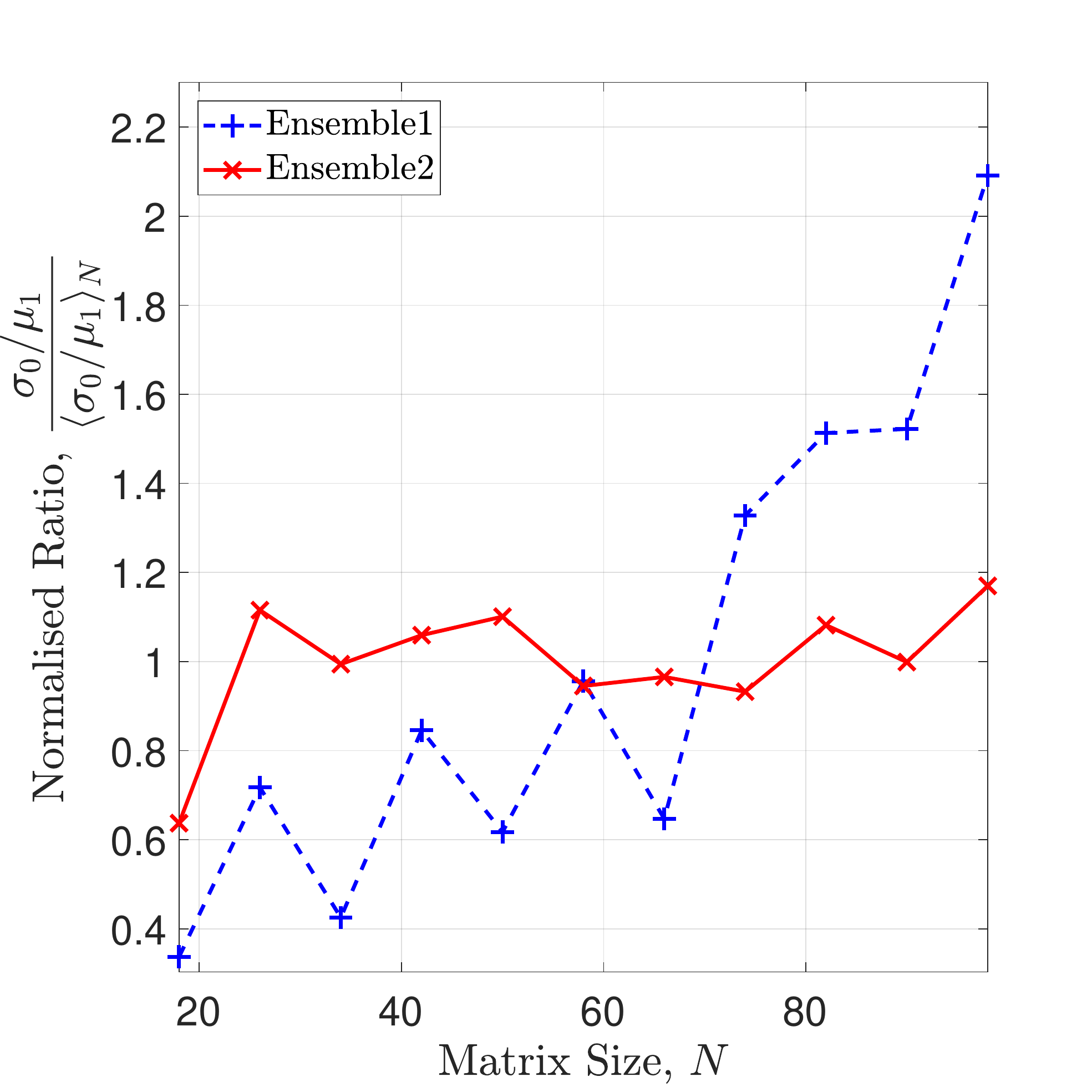}
		\includegraphics[width=0.49\linewidth,angle=0]{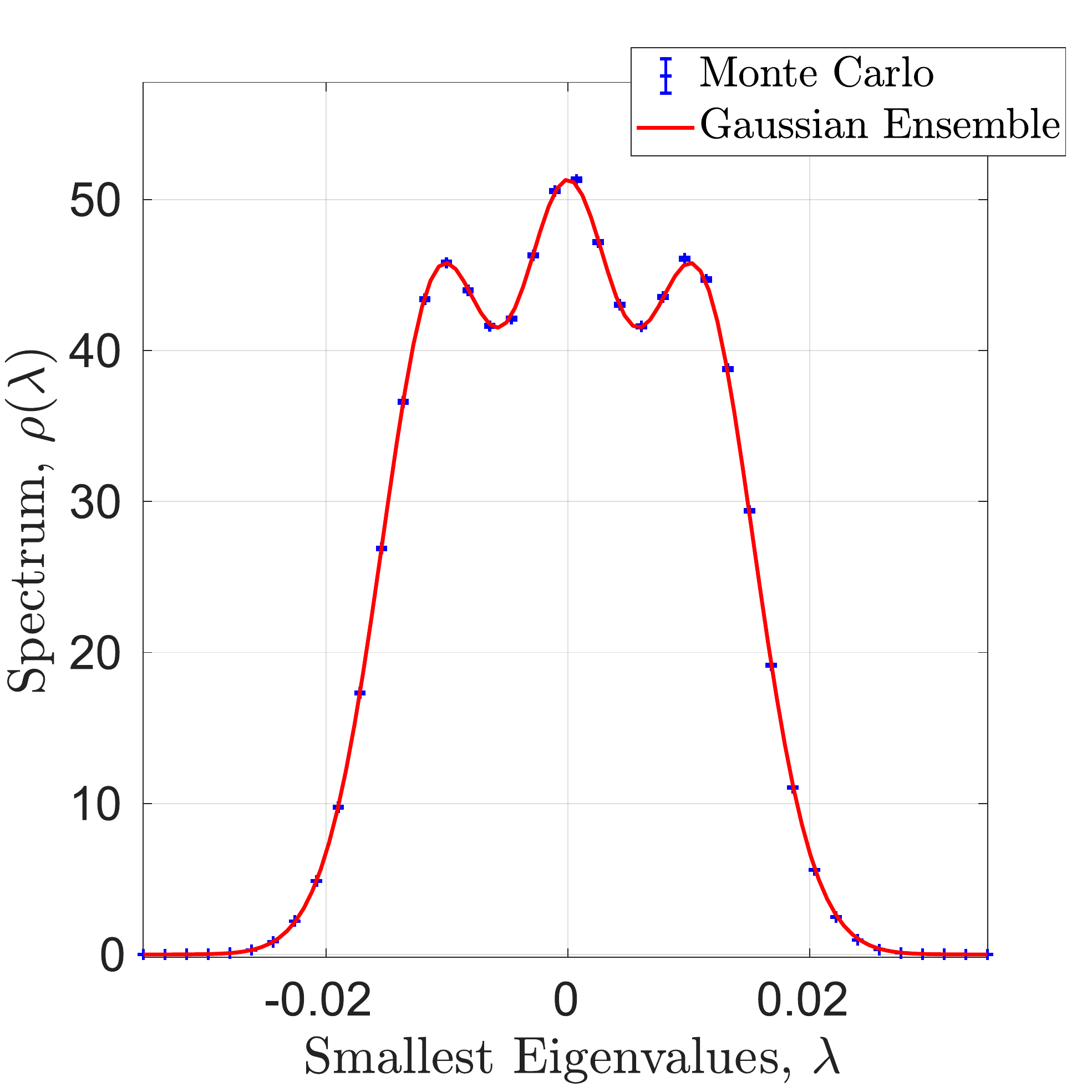}
	\caption{
		\textbf{(a)} Identification of former topological modes: A comparison of the ratio between the width of the smallest eigenvalue and the position of the second smallest eigenvalue as a function of the matrix size $N$ for \EnsAntiRand\ ($\nu=1$) and \EnsAntiNone\ ($\nu=0$) in Section \ref{Sec:Example}. We have normalised the mean of each curve. The coupling constant is set to $\alpha= 0.01 \sqrt{\Tr A^{-2}}||S||_{\rm op}/\sqrt{N}$ according to \eqref{Eq:strength}. The ensemble size is $10^{5}$.
		\textbf{(b)} The density of the smallest eigenvalues for \EnsCh\ with $n = 33$ and $\nu=3$ from Section \ref{Sec:Example}. The Monte Carlo simulation (blue error bars, $10^{6}$ matrices generated) and the theoretical distribution of the GUE of size $3\times 3$ (red solid curve) are compared, see \eqref{Eq:rhoGUE}.
	}
	\label{Fig:Scaling}
\end{figure}

Following \eqref{heuristic:alpha.4}, the width of the former zero eigenvalues is $\alpha\sqrt{\Tr S^2}/N$ and the smallest eigenvalues of A are $1/N$. Rescaling of the eigenvalues according to \eqref{Eq:EigRescale} yields a broadening of $\alpha\sqrt{\Tr S^2}$. Assuming $\Tr S^2\sim N$ and fixed $\alpha$, the width of the rescaled broadened zero modes scale as $\sqrt{N}$. We will demonstrate how this different scaling can be used as an experimental identifier of topological modes. We also illustrate this in Figure \ref{Fig:Scaling} (a).

\subsection{Application to Experiments}\label{Sec:ApplicationsExp}
We wish to relate the scaling with $N$ to physical quantities.
We here use a result from the $\epsilon$-regime of effective field theory, namely that the size of the matrix scales linearly with the volume of the system. We refer to \cite{RMT_2_EFT-1,RMT_2_EFT-2} for the full derivation, but the general idea is to calculate the non-linear $\sigma$-model (or chiral Lagrangian) of the random matrix model, which for all classes has the form
\begin{eqnarray}
S &=& \int d^4x\left[\Tr \left(\partial_\mu U\partial^\mu U\inv\right)+\Tr M(U+U\inv)\right] .
\end{eqnarray} The exact nature of the Goldstone field $U$ will depend on the class. As we consider the low-energy modes around the origin, where dynamics are frozen out \cite{DOTV,GasserLeutwylerThermo,GasserLeutwylerSym}, the potential term becomes the most important. Constructing the Lagrangian directly from the matrix model leads to the identification $V\sim N$.
This means that under the above assumptions, the width of the broadened modes scale as $\sqrt{V}$.

The proposed identifier is therefore the ratio $\sigma_0/\mu_1$, where $\sigma_0$ is the width of the ground state distribution, and $\mu_1$ is the mean position of the first excited state. If this scales significantly different from 1, it is safe to conclude a system with a broadened zero mode.
This scaling is also found in the literature of lattice QCD and has helped to explain the unusual behaviour observed in lattice simulations \cite{DSV,ADSV}.

\subsection{Example Ensembles}\label{Sec:Example}
For the numerical checks, we compare the following four ensembles. We first draw a fixed $A^{(N)}$ and $S^{(N)}$ and then we average over $U$ for the Hamiltonian $K^{(N)}=A^{(N)}+\alpha US^{(N)}U\hc$.

\EnsAntiRand\textit{:}
To illustrate a particular condensed matter application we consider a direct sum of two antisymmetric matrices that are the same up to a sign, corresponding to particle-hole-symmetry \cite{Chiu,BeenakkerRMT,Neven}. This ensemble is perturbed by off-diagonal blocks in order to model topological superconductors carrying Majorana modes. The ensemble has the form
\begin{eqnarray}\label{Eq:EnsAntiRand}
K^{(N)} &=& \left(\begin{matrix}
iM & 0\\
0 & -iM
\end{matrix}\right) + \alpha O\left(\begin{matrix}
0 & iW\\
-iW^T & 0
\end{matrix}\right)O^T\ ,\ M = -M^T\ .
\end{eqnarray}
The matrices $M$ and $W$ are real and of dimension $2n+\nu$, and $M$ is antisymmetric. So for $\alpha=0$ and $\nu = 1$ the model exhibits two generic zero modes. The matrices are generated once via i.i.d.~entries uniform on the interval $[-1,1]$ and then kept fixed. The ensemble average is only done via the orthogonal matrix $O$. The full matrix $K^{(N)}$ is of size $N = 4n+2\nu$ and imaginary antisymmetric, and for $\alpha>0$ no exact modes are present.
For $\nu=1$ the two zero modes are broadened by the coupling. They form a $2\times 2$ imaginary antisymmetric Gaussian ensemble.

\EnsAntiNone\textit{:} To illustrate the different scalings of broadened zero eigenvalues and bulk eigenvalues, we also consider an ensemble for comparison of the form
\begin{eqnarray}\label{Eq:EnsAntiNone}
K^{(N)} = iA^{(N)} + i\alpha OS^{(N)}O^T\ ,\ 
K^{(N)} = {K^{(N)}}\hc = -{K^{(N)}}^T
\end{eqnarray}
with matrix size $N=2n$ and no further substructure. This ensemble never has exact zero modes in contrast to the models covered by our discussion. We again draw all matrix entries of $A$ and $S$ i.i.d.~once, uniformly from the interval $[-1,1]$. Afterwards we keep them fixed and average over the orthogonal matrices $O$ only.

\begin{figure}
	\hspace{-0.39\linewidth}	\textbf{(a)}\hspace{0.47\linewidth}	\textbf{(b)}\hfill\\
	\includegraphics[width=0.49\linewidth,angle=0]{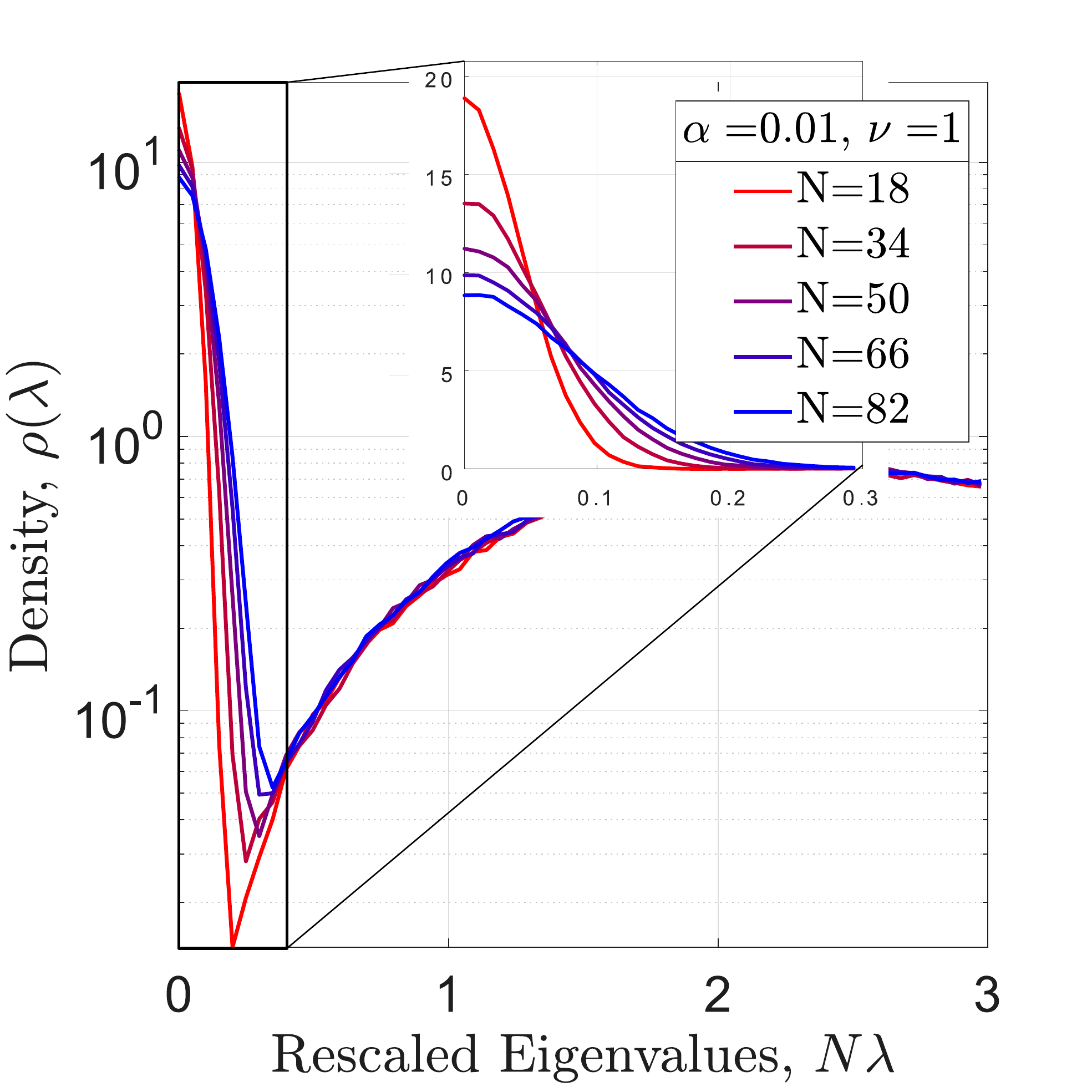} \includegraphics[width=0.49\linewidth,angle=0]{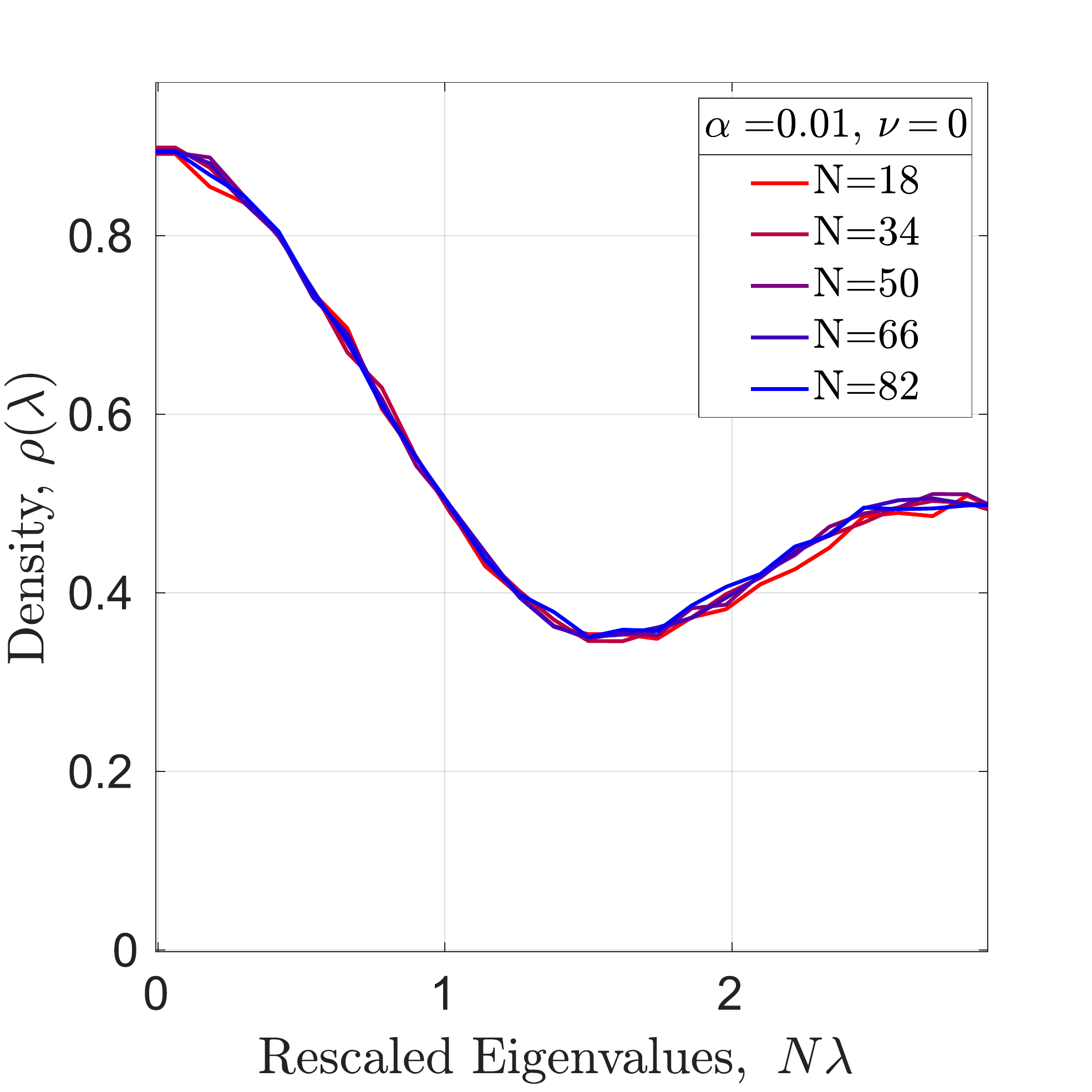}
	\caption{
		The microscopic density for Monte Carlo simulations of an ensemble with a single topological mode and one without (see \ref{Eq:EnsAntiRand} and \ref{Eq:EnsAntiNone}) for different matrix sizes. Here we have also averaged over the spectrum for visual clarity. The eigenvalues have been rescaled according to \eqref{Eq:EigRescale} to keep the distance between the smallest eigenvalues of the order $1$. We compare the difference between a topological and a non-topological mode. We see the former topological mode broaden with $N$. The ensemble size is $10^{5}$ and the bin size is roughly $0.2$ for (a) plot and $0.1$ for (b). The density in (a) is shown on logarithmic scale to keep both peaks visible in the same plot, but a zoom-in is provided.
	}
	\label{Fig:Density}
\end{figure}

In Figure \ref{Fig:Density} we compare the microscopic densities about the origin for both \textit{Ensembles 1} and \textit{2}. In both plots we have rescaled the eigenvalues according to \eqref{Eq:EigRescale} to keep the mean inter-eigenvalue distance of order 1. We have also averaged over the spectrum of $A$ and $S$, which was not the case in Figure \ref{Fig:Scaling} (a). This is done to increase the contrast of the scaling of the eigenvalues with the volume $V$ represented by $N$.
As predicted in Section \ref{Sec:ApplicationsExp}, the perturbed zero mode in \EnsAntiRand\ changes with the volume in the rescaled variables, whereas the same does not happen for the smallest eigenvalue in \EnsAntiNone.

However, averaging over the spectrum is not necessary as we show in Figure \ref{Fig:Scaling} (a), where we plot the ratio $\sigma_0/\mu_1$ as a function of the matrix size $N$. We suggest this quantity as an identifier for topological or other system specific zero modes. We rescale $\alpha ||S||_{\rm op}\sqrt{\Tr A^{-2}}/\sqrt{N}\to \alpha$ to keep the coupling constant on the same scale for all matrix sizes, see \eqref{Eq:strength}. As we do not average over the spectrum, the variance of the individual modes partially obscures the scaling, but it is still visible. If an average over the spectrum is also performed, the difference becomes even clearer, cf. Figure \ref{Fig:Density}.

\begin{figure}
\hspace{-0.39\linewidth}	\textbf{(a)}\hspace{0.47\linewidth}	\textbf{(b)}\hfill\\
		\centering
		\includegraphics[width=0.49\linewidth,angle=0]{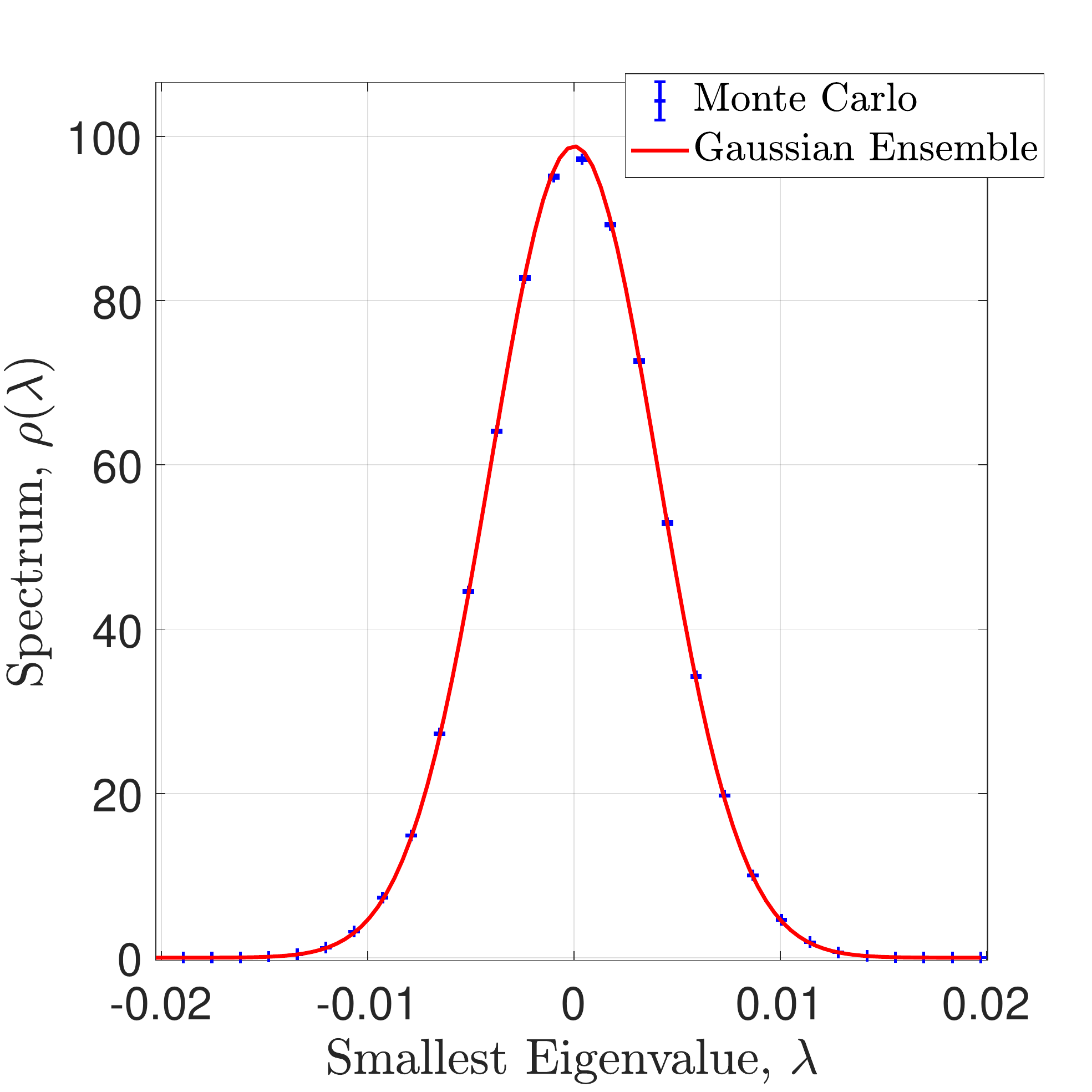}
		\includegraphics[width=0.49\linewidth,angle=0]{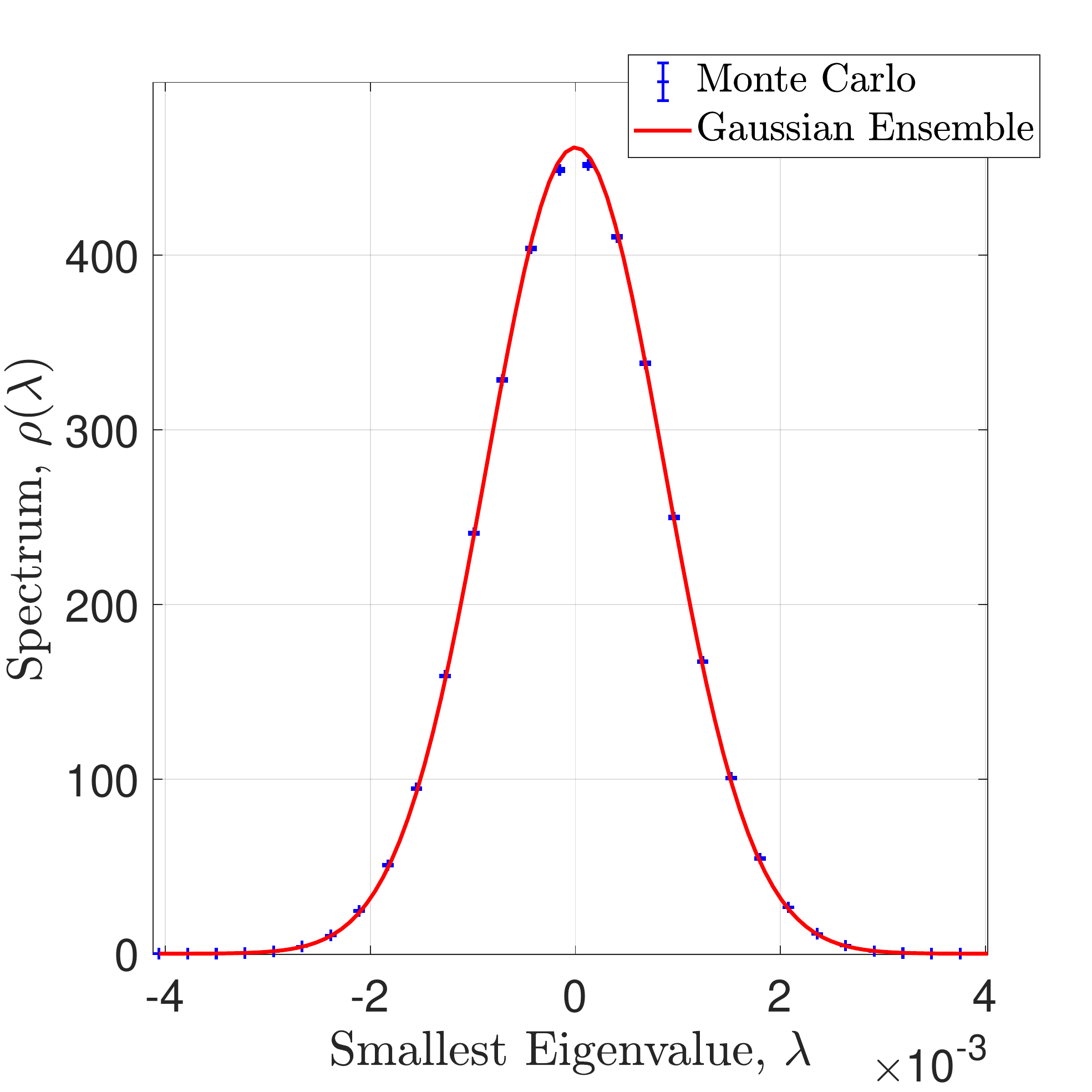}		
	
	\caption{
		Distribution of the two smallest eigenvalues of \EnsAntiRand\ (a) and \EnsAntiGam\ (b) in Section \ref{Sec:Example}. In both ensembles the unperturbed Hamiltonian $A$ is imaginary, antisymmetric and block-diagonal of dimension $N=134$ so that it has two zero eigenvalues. The perturbation $S$ is a full generic imaginary matrix $i \alpha W$ (\EnsAntiRand) on the off-diagonal block and a constant matrix $i\alpha \eins$ (\EnsAntiGam) with $\alpha=0.01$. The Monte Carlo simulations (blue error bars, $10^{6}$ matrices generated) are compared with our theoretical RMT predictions that are Gaussian distributions with the correct variances derived in Section \ref{Sec:Universality} (red solid curves).
	}
	
	\label{Fig:Compare}
\end{figure}

\EnsAntiGam\textit{:} To illustrate that degeneracy of the perturbation is irrelevant as long as it satisfies the conditions \CondS, we consider an ensemble very similar to \EnsAntiRand, except that the perturbation is proportional to the second Pauli matrix. That is,
\begin{eqnarray}\label{Eq:EnsAntiGam}
K^{(N)} &=& \left(\begin{matrix}
iM_1 & 0\\
0 & iM_2
\end{matrix}\right) + i\alpha O\left(\begin{matrix}
0 & \eins\\
-\eins & 0
\end{matrix}\right)O^T\ ,\\
&& M_1 = -{M_1}^T\ ,\ M_2 = -{M_2}^T\ .\nn
\end{eqnarray}
$M_1$ and $M_2$ are real antisymmetric, but independent as the eigenvalues would otherwise be shifted rather than perturbed. These are chosen fixed with i.i.d.~entries on the interval $[-1,1]$ while the average is over the orthogonal matrix $O$. With this ensemble we would like to emphasise the generality of the conditions \CondS. That is, the matrix Central Limit Theorem stated above describes the limit for a broad class of ensembles.
This similarity is illustrated in Figure \ref{Fig:Compare} where we compare Monte Carlo simulations to the corresponding theoretical curves derived in Section \ref{Sec:Universality}.

\EnsCh\textit{:} As an application to QCD, more precisely lattice QCD, where chirality is broken by a
perturbation \cite{DWW2011,DHS2012,KieburgWilson,DSV,ADSV,MarioJacWilson}, we consider the following model
\begin{eqnarray}\label{Eq:EnsCh}
K^{(N)} &=& \left(\begin{matrix}
0 & M\\
M\hc & 0
\end{matrix}\right) + \alpha USU\hc\ .
\end{eqnarray}
$M$ is a complex $(n+\nu)\times n$ matrix with no further symmetries, $S$ is a complex hermitian matrix, and $U$ is unitary and Haar-distributed. As before the only average we perform is over $U$. The index $\nu$ determines the number of exact zero modes, which allows us to have any number of broadened modes, unlike the antisymmetric ensembles. The $\nu$ zero modes from the chiral ensemble are all broadened by the perturbation, which is hermitian and has no further symmetry. This means that the former zero modes are distributed according to a Gaussian unitary ensemble of size $\nu\times \nu$ \cite{Mehta}
\begin{eqnarray}\label{Eq:rhoGUE}
\rho^\nu_{\rm GUE}(\lambda) &=& \frac{1}{2\sigma}\sum_{j=0}^{\nu-1} \varphi_j\left(\frac{\lambda}{\sigma}\right)^2\ ,\\
\varphi_j(\lambda) &=& \frac{1}{\sqrt{2^j j! \sqrt{\pi}}} e^{-\lambda^2/2} H_j(\lambda)\nn
\end{eqnarray}
with $\sigma = \alpha\sqrt{\Tr  (S^{(N)})^2/(\gamma N^2)}$ the Hermite polynomials corresponding to the weight $e^{-\lambda^2}$.
In Figure \ref{Fig:Scaling} (b) we compare the broadening of this ensemble to the theoretical prediction with the width found in Section \ref{Sec:CondUn1}.

\section{Conclusion} \label{Sec:Conclusion}

We have presented a general mechanism explaining the observation of the universal broadening of degenerate eigenvalues inside a spectral gap when a generic perturbation is switched on. This universality states that the broadening follows the statistics of a finite-dimensional Gaussian random matrix ensemble. Exactly the finite dimensionality is surprising because one usually expects that spectral universality only holds in the limit of large matrix dimensions. This new universality relies on a self-average of the change of basis $U_2=\{\langle\psi_j|\phi_l\rangle\}_{j,l= N - \nu + 1,\dots,N}$ between the unperturbed operator $A$ and the perturbation $S$ associated with the zero modes of $A$. In the present work, we have averaged over all bases transformations $U_2$ drawn from the Haar measure of the group associated to the respective symmetry class. Yet lattice simulations in QCD~\cite{DelDebbio:2005qa,DWW2011,DHS2012,CGRSZ} strongly suggest that the measure can be relaxed to something non-uniform. As a further study it is natural to investigate what happens if the assumption of an average over the full Haar measure is loosened.

In our analysis, we quantified the conditions under which this universal broadening holds. The three conditions~(\ref{Eq:CondTrace0}-\ref{Eq:strength}) are rather mild and have very natural physical interpretations like the relation between closing of the spectral gap and the coupling strength $\alpha$. Especially, we recover the critical scaling of $\alpha$ found in lattice QCD with Wilson fermions~\cite{DelDebbio:2005qa,DWW2011,DHS2012,CGRSZ} and in the RMT-models for Majorana modes in disordered quantum wires~\cite{AKMV,KimAdam}.

As a possible application we have suggested that our results may be used to distinguish topological modes in the bulk from modes in the bulk. The scaling behaviour in the system size and the coupling parameter $\alpha$ of the broadening for the eigenvalues of the two kind of modes is completely different. Consequently, this scaling might provide an ideal indicator of experiments.

{\it Acknowledgements:} 

We would like to thank J.~J.~M.~Verbaarschot and G.~Akemann for interesting discussions on the subject. The first idea for the symmetry breaking in topological superconductors was conceived in collaboration with P.~H.~Damgaard, K.~Flensberg, and E.~B.~Hansen. K.~S.~would like to thank A. Altland for discussions.
Support by the German research council (DFG) through CRC 1283: ``Taming uncertainty and profiting from randomness and low regularity in analysis, stochastics and their applications"  (M.~K.) and
International Research Training Group 2235 Bielefeld-Seoul "Searching for the regular in the irregular: Analysis of singular and random systems" (A.~M.) is kindly acknowledged.
A.~M.~would also like to thank Stony Brook University for their hospitality in October 2017 and "Bielefeld Graduate School in Theoretical Sciences Mobility Grant" for funding the stay.

{\it E-mail address:}

\noindent M.~Kieburg: \url{mkieburg@physik.uni-bielefeld.de}\\
A.~Mielke: \url{amielke@math.uni-bielefeld.de}\\
K.~Splittorff: \url{split@nbi.ku.dk}

\end{document}